\documentclass[aps,preprint,superscriptaddress]{revtex4-1}

\usepackage{amsmath}    
\usepackage{graphicx}   
\usepackage{color}      
\usepackage{subfigure}  
\usepackage{hyperref}   
\usepackage{todonotes}
\usepackage{nicefrac}
\usepackage[utf8]{inputenc}

\begin{document}

\title{Plasmons and Interband Transitions of Ca$_{11}$Sr$_3$Cu$_{24}$O$_{41}$ investigated by Electron Energy-Loss Spectroscopy}
\author{Friedrich Roth}
\author{Christian Hess}
\author{Bernd B\"uchner}
\affiliation{IFW Dresden, P.O. Box 270116, D-01171 Dresden, Germany}
\author{Udo Ammerahl}
\author{Alexandre Revcolevschi}
\affiliation{Laboratoire de Physico-Chimie de l’État Solide, Université Paris-Sud, 91405 Orsay, France}
\author{Martin Knupfer}
\affiliation{IFW Dresden, P.O. Box 270116, D-01171 Dresden, Germany}
\date{\today}

\begin{abstract}
Electron energy-loss spectroscopy studies have been performed in order to get a deeper insight into the electronic structure and elementary
excitations of the two\,-\,leg ladder system Ca$_{11}$Sr$_3$Cu$_{24}$O$_{41}$. We find a strong anisotropy of the loss function for momentum
transfers along the $a$ and $c$\,-\,crystallographic axis, and a remarkable linear plasmon dispersion for a momentum transfer parallel to the
legs of the ladders. The investigated spectral features are attributed to localized and delocalized charge-transfer excitations and the charge
carrier plasmon. The charge carrier plasmon position and dispersion in the long wave-length limit agree well with expectations based upon the
band structure of the two-leg ladder, while the observed quasi-linear plasmon dispersion might be related to the peculiar properties of
underdoped cuprates in general.
\end{abstract}

 \maketitle

\section{Introduction}
\begin{figure}[h]
\includegraphics[width=0.95\textwidth]{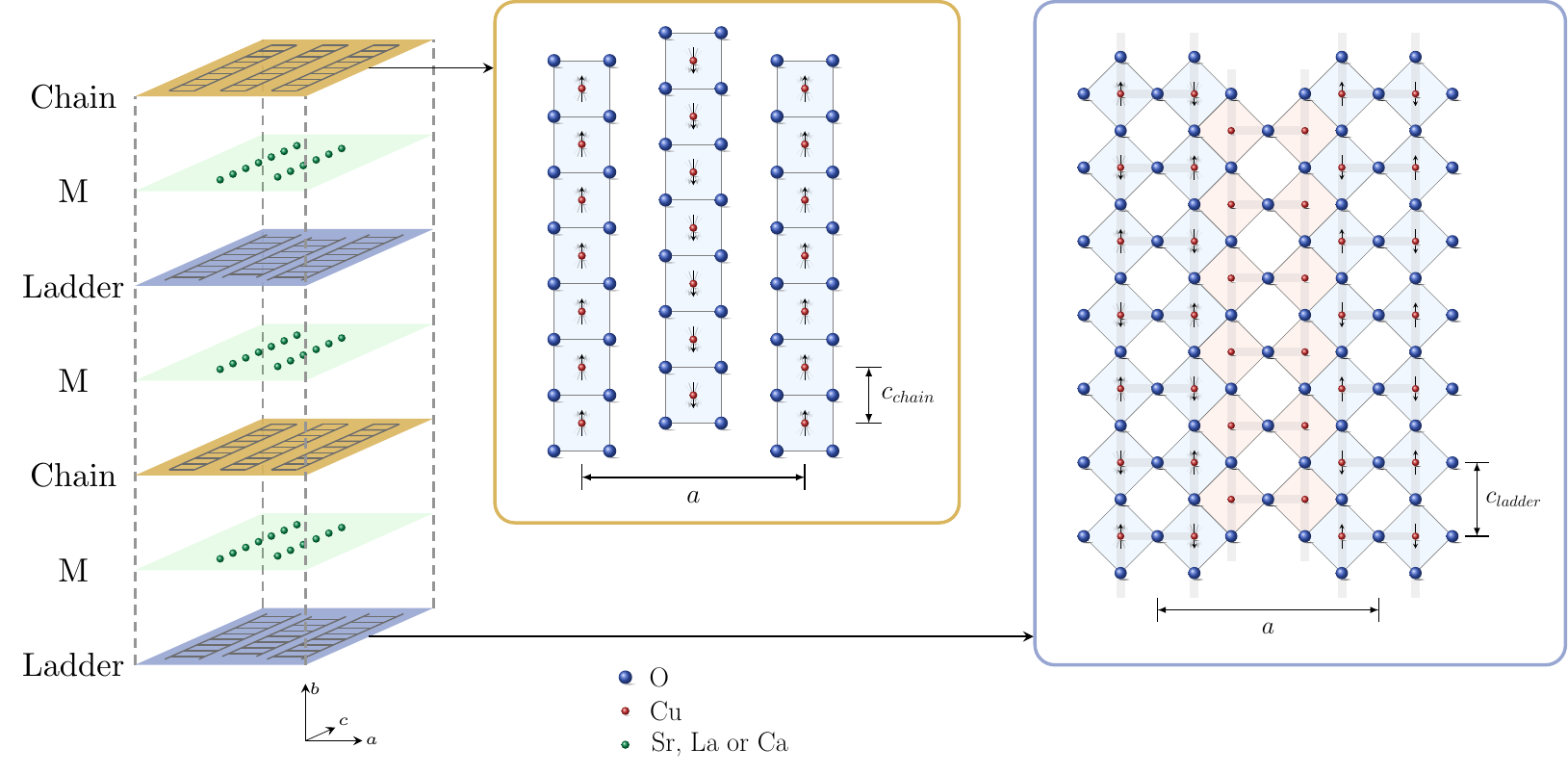}
\caption{\label{fig1}Schematic representation of the crystall structure of Ca$_{x}$Sr$_{14-x}$Cu$_{24}$O$_{41}$}
\end{figure}

\noindent Since the discovery of high-temperature
superconductivity in Cu-O frameworks \cite{Bednorz1986} a large
family of different Cu-O based systems was studied, whereas the
dimensionality varied from quasi zero-dimensional systems (like
Li$_2$CuO$_2$ or NaCuO$_2$) over one-dimensional networks (e.\,g.
Sr$_2$CuO$_3$) to two-dimensional systems (such as
Sr$_2$CuO$_2$Cl$_2$ or high temperature superconductors). The
dimension of a system and the associated electronic and magnetic
pathway joining neighboring Cu ions, which depends upon the manner
in which the CuO$_4$ plaquettes are arranged, plays a key role for
the electronic excitations. The compound
Ca$_{x}$Sr$_{14-x}$Cu$_{24}$O$_{41}$ is a so-called
quasi-one-dimensional system and shows additional complexity since
it consists of two different types of copper oxide networks---CuO$_2$ (edge-sharing) chains and  Cu$_2$O$_3$ (two-leg)
ladders---which are separated by strings of Sr or Ca atoms. These
networks are arranged in layers, and the layers are oriented in the
crystallographic $ac$-plane, while they are stacked in an
alternating manner along the perpendicular $b$-axis (see Fig.\,\ref{fig1}). Both of these two subsystems, ladders and chains,
have orthorhombic symmetry, but are structurally
incommensurate.\cite{McCarron1998,Siegrist1988} The discovery of
superconductivity in Ca$_{13.6}$Sr$_{0.4}$Cu$_{24}$O$_{41}$ at a
high pressure of 3\,GPa \cite{Uehara1996} has provoked a lot of
attention on the spin-ladder system because it was the first
superconducting copper oxide material with a
non-square-lattice.\cite{Nagata1997} A remarkable property of
Ca$_{x}$Sr$_{14-x}$Cu$_{24}$O$_{41}$ is that superconductivity
only occurs when Sr is replaced by Ca. Thereby, the nominal
valence of Cu remains unchanged but the change of chemical
pressure within the lattice causes a transfer of holes from the
chain to the ladder
subsystem.\cite{Kato1996,Nuecker2000,Koitzsch2010} Therefore, the
evolution of the electronic and magnetic structure upon Ca addition is one of
the key issues for understanding superconductivity and other
physical properties \cite{Kataev2001,Ammerahl2000}, whereas the exact hole distribution in
these compounds is still under debate.\cite{Nuecker2000,Osafune1997,Rusydi2007,Magishi1998} 

\noindent Another interesting property of the spin ladder compounds is a tendency to form a charge density wave (CDW) phase depending on the Ca
content \cite{Blumberg2002,Vuletic2003,Hess2004}, which may prevent the occurrence of superconductivity. An insulating hole crystal phase, as it was
predicted \cite{Dagotto1992}, in which the carriers are localized through manybody interactions was
reported.\cite{Abbamonte2004,Carr2002,Friedel2002} In summary, the complex phase diagram as well as the effect of dimensionality and the impact
of temperature on the electronic structure of these compounds are not yet fully understood.

Electron energy-loss spectroscopy (EELS) is a useful tool for the
investigation of materials at all levels of complexity in the
electronic-spectrum.\cite{Fink2001} The EELS cross section is
basically proportional to
$\operatorname{Im}$[-1/$\epsilon(\omega,\bf q)$] (called loss
function) where $\epsilon(\omega,\bf q)$ = $\epsilon_1(\omega,\bf
q)$ + $i \epsilon_2(\omega,\bf q)$ is the momentum and
energy-dependent complex dielectric function. In this way, EELS
probes the electronic excitations of a solid  under investigation.
Furthermore, it allows momentum dependent measurements of the loss
function, i.\,e. the observation of non-vertical transitions within
the band structure of a solid, the idendification of dipole forbidden excitations \cite{Knupfer1999,
Atzkern2000} and the determination of the dispersion of excitons, interband transitions or charge carrier
plasmons.\cite{Nuecker1989,Wang1990,Nuecker1991,Romberg1990,Schuster2007}
As the dispersion of a charge carrier plasmon is related to the
Fermi velocity, EELS studies also provide valuable
insights into further fundamental electronic parameters.

\section{Experimental}
\noindent Single crystals of Ca$_{11}$Sr$_3$Cu$_{24}$O$_{41}$ were grown by using the travelling solvent floating zone
method.\cite{Ammerahl1998} For the EELS measurements thin films ($\sim$ 100\,nm) were cut along the crystal $b$-axis from these single crystals
using an ultramicrotome equipped with a diamond knife. The films were then put onto standard transmission electron microscopy grids and
transferred into the spectrometer. All measurements were carried out at room temperature with a dedicated transmission electron energy-loss
spectrometer \cite{Fink1989} employing a primary electron energy of 172\,keV. The energy and momentum resolution were set to be $\Delta
E$\,=\,80\,meV and $\Delta q$\,=\,0.035\,\AA$^{-1}$, respectively. Before measuring the loss-function, the thin films have been
characterized by \textit{in situ} electron diffraction, in order to orient the crystallographic axis with respect to the
transferred momentum. From the measured loss function, the real and imaginary part of the dielectric function $\epsilon(\omega)$ and
consequently the optical conductivity $\sigma$ were calculated by the well-known Kramers-Kronig relations.\cite{Fink1989} Prior to the
Kramers-Kronig analysis the measured spectra were corrected by substracting contributions of multiple scattering and eliminating the
contribution of the direct beam by fitting the plasmon peak with a model function, which gives the low energy tail to zero energy for the loss function.\cite{Nuecker1989} We note that in our case the quasi elastic background does not alter the plasmon position as discribed in \cite{Chen1977,Batson1983,Bertoni2010}.

\section{Results and Discussion}

\begin{figure}[h]
\includegraphics[width=0.7\textwidth]{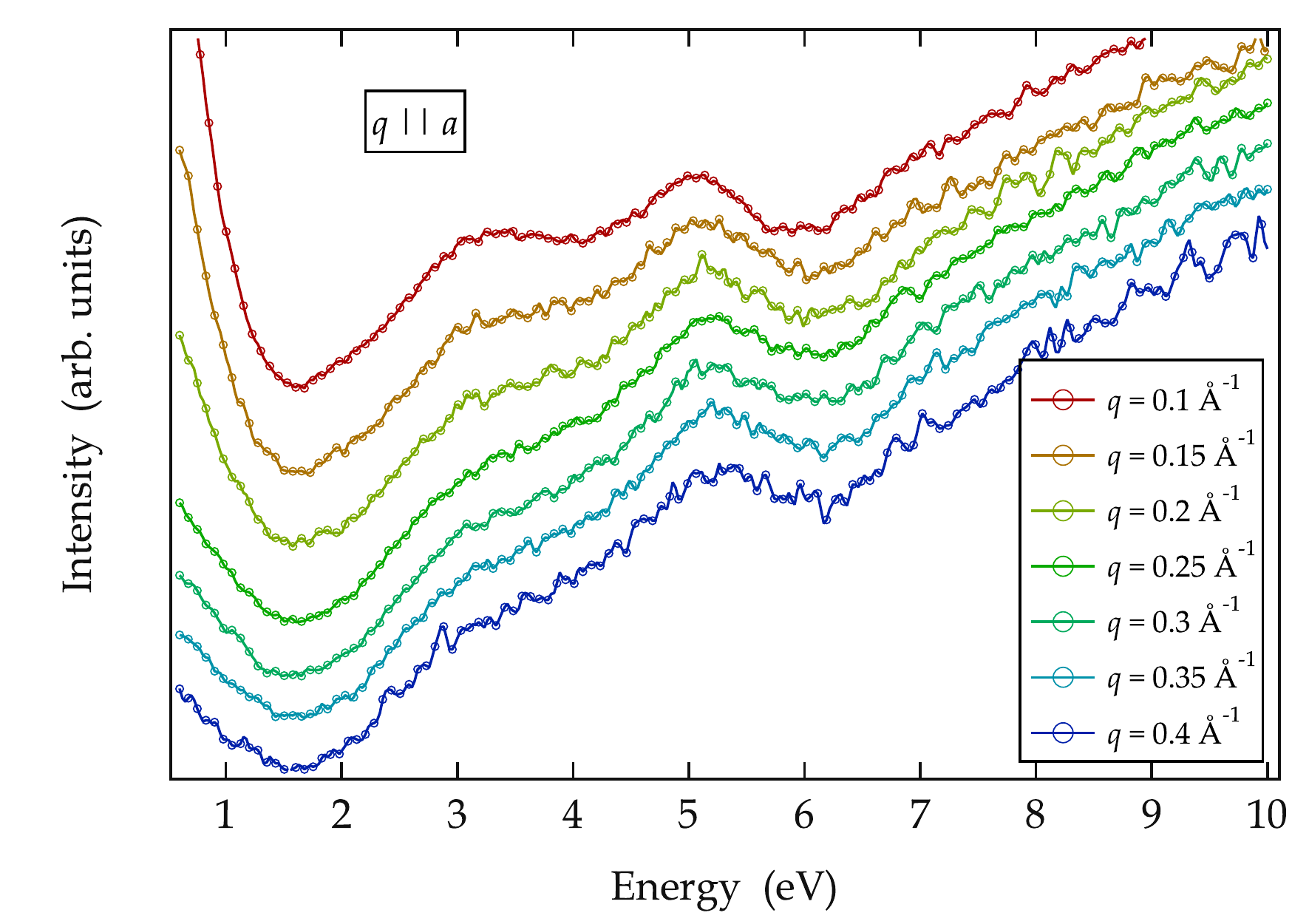}
\caption{\label{fig2}The momentum dependence of the EELS spectra
of Ca$_{11}$Sr$_3$Cu$_{24}$O$_{41}$ for $q$ parallel to the
crystallographic $a$\,-\,axis ($q$ is increasing from top to bottom spectra). The upturn towards 0\,eV is due to the quasi-elastic line.}
\end{figure}

\begin{figure}[h]
\includegraphics[width=0.7\textwidth]{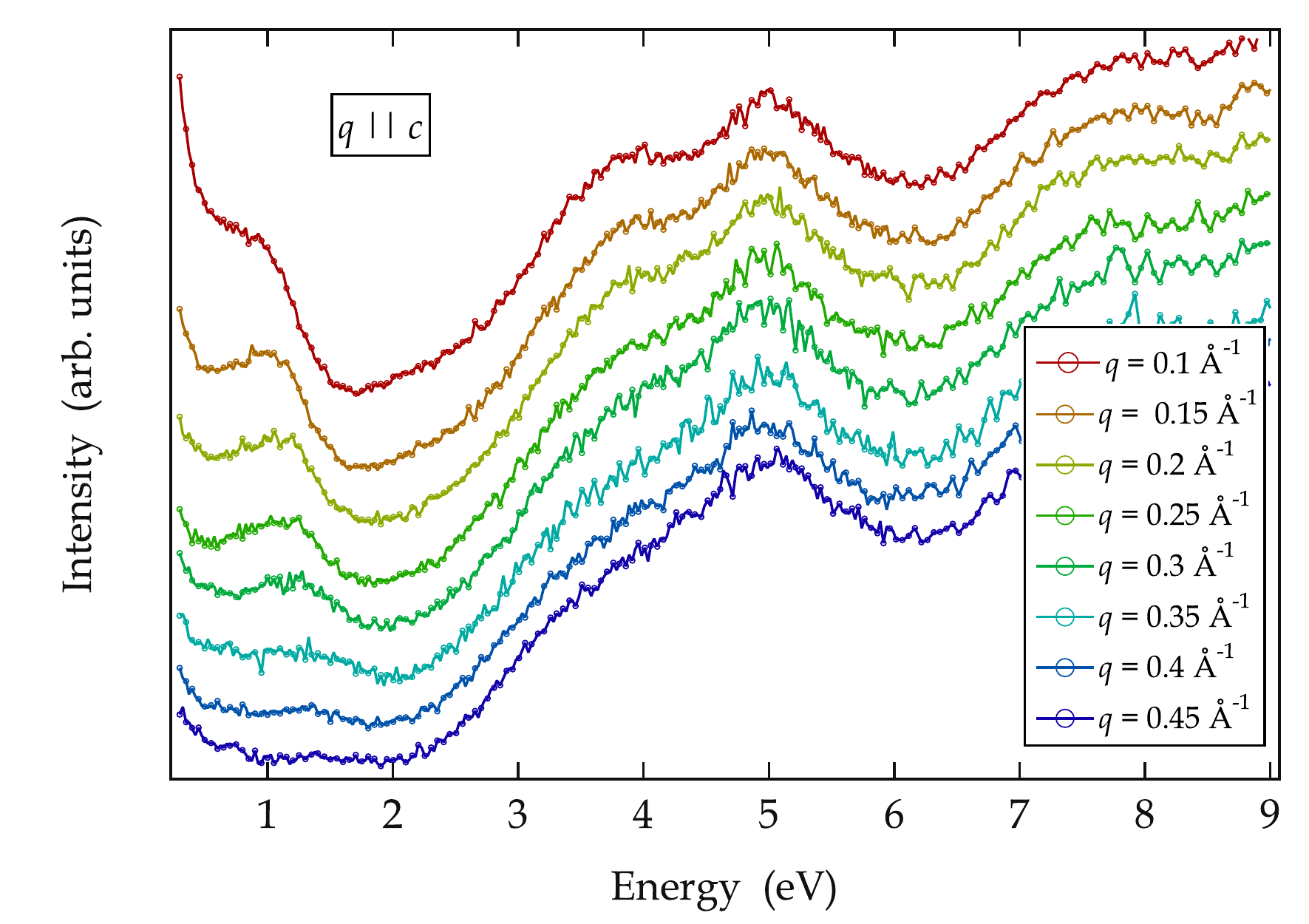}
\caption{\label{fig3}The momentum dependence of the EELS spectra
of Ca$_{11}$Sr$_3$Cu$_{24}$O$_{41}$ for $q$ parallel to the
crystallographic $c$\,-\,axis ($q$ is increasing from top to bottom spectra). The upturn towards 0\,eV is due to the quasi-elastic line.}
\end{figure}

\noindent In Fig.\,\ref{fig2} we show the evolution of the loss function with increasing $q$ in an energy range between 0.5\,-\,10\,eV for a
momentum transfer perpendicular to the ladders/chains (crystallographic $a$\,-\,axis). The data have been normalized to the high-energy region
between 9 and 10\,eV where they are almost momentum independent. We can clearly identify a well pronounced double peak structure with maxima at
3\,-\,3.5\,eV and at 5\,eV. The spectral weight of the first excitation feature---compared  to the second---decreases with increasing momentum
transfer. Fig.\,\ref{fig3} displays the corresponding data for a momentum transfer parallel to the crystallographic $c$\,-\,axis, i.\,e. parallel
to the ladders and chains in Ca$_{11}$Sr$_3$Cu$_{24}$O$_{41}$. Again, there is a double-peak feature between 3 and 5 eV, and the intensity of
the former is decreasing upon increasing momentum.

\noindent In addition, Fig.\,\ref{fig3} reveals a further excitation feature around 1 eV for momentum transfers parallel to the ladder
direction, which is absent perpendicular to it. This additional excitation clearly disperses to higher energies with increasing $q$.
Furthermore, the peak width increases with increasing momentum, which indicates damping of this excitation which also increases with $q$.
According to resistivity data \cite{Motoyama1997}, Ca$_{11}$Sr$_3$Cu$_{24}$O$_{41}$ shows a metallic behavior along  the $c$\,-\,direction,
which is also in line with the appearance of a plasma edge close to 1\,eV in the corresponding reflectivity spectra.\cite{Osafune1997,
Ruzicka1998} Consequently, we ascribe the peak around 1\,eV to the so\,-\,called Drude plasmon (or charge carrier plasmon) caused by the
collective excitation of the free charge carriers. This is analogous to what has been observed for other doped cuprate
systems.\cite{Wang1990,Nuecker1991,Knupfer1994}

\begin{figure}[h]
\includegraphics[width=0.7\textwidth]{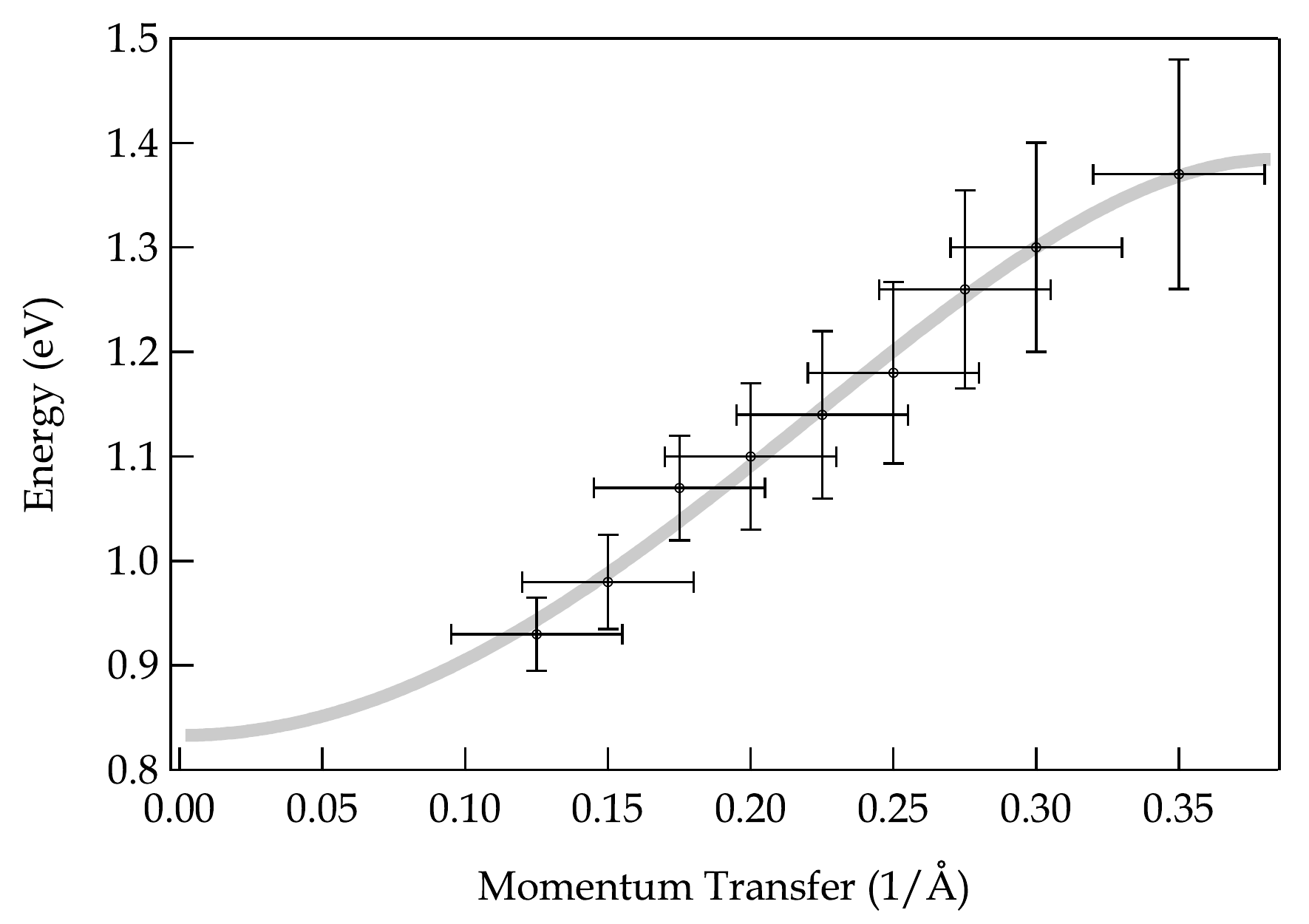}
\caption{\label{fig4}Plasmon dispersion in Ca$_{11}$Sr$_3$Cu$_{24}$O$_{41}$ along the $c$\,-\,direction. Within the error bars the plasmon scales linearly with momentum and the bandwidth amounts to $\approx$ 400\,meV in the considered momentum range. The gray curve represent the fit with a polynomial function (cf. equation (2))}
\end{figure}

\noindent In order to further quantify the behavior of the 1\,eV plasmon we present in Fig.\,\ref{fig4} the evolution of the peak
position in the range 0.15\,\AA$^{-1}$ to 0.35\,\AA$^{-1}$ for $q
\parallel$ $c$. Due to the strong damping of the
plasmon and the low cross section for higher momentum transfers,
data for a momentum transfer above $q$\,=\,0.35\,\AA$^{-1}$ are
not included in Fig.\,\ref{fig4}. It can be seen that the plasmon
dispersion in Ca$_{11}$Sr$_3$Cu$_{24}$O$_{41}$ is positive with a
bandwidth of at least 400\,meV. This is in very good
agreement with the dispersion found in planar cuprates such as
Bi$_2$Sr$_2$CaCu$_2$O$_{8-\delta}$.\cite{Nuecker1991} Moreover,
the plasmon dispersion is linear in $q$, which is in contrast to
what one would expect for an ``ordinary'' metallic plasmon, where
it should be quadratic.\cite{Sturm1982,Pines1963} We note that for
Bi$_2$Sr$_2$CaCu$_2$O$_{8-\delta}$ a quadratic plasmon dispersion
has been reported.

\begin{figure}
\includegraphics[width=0.7\textwidth]{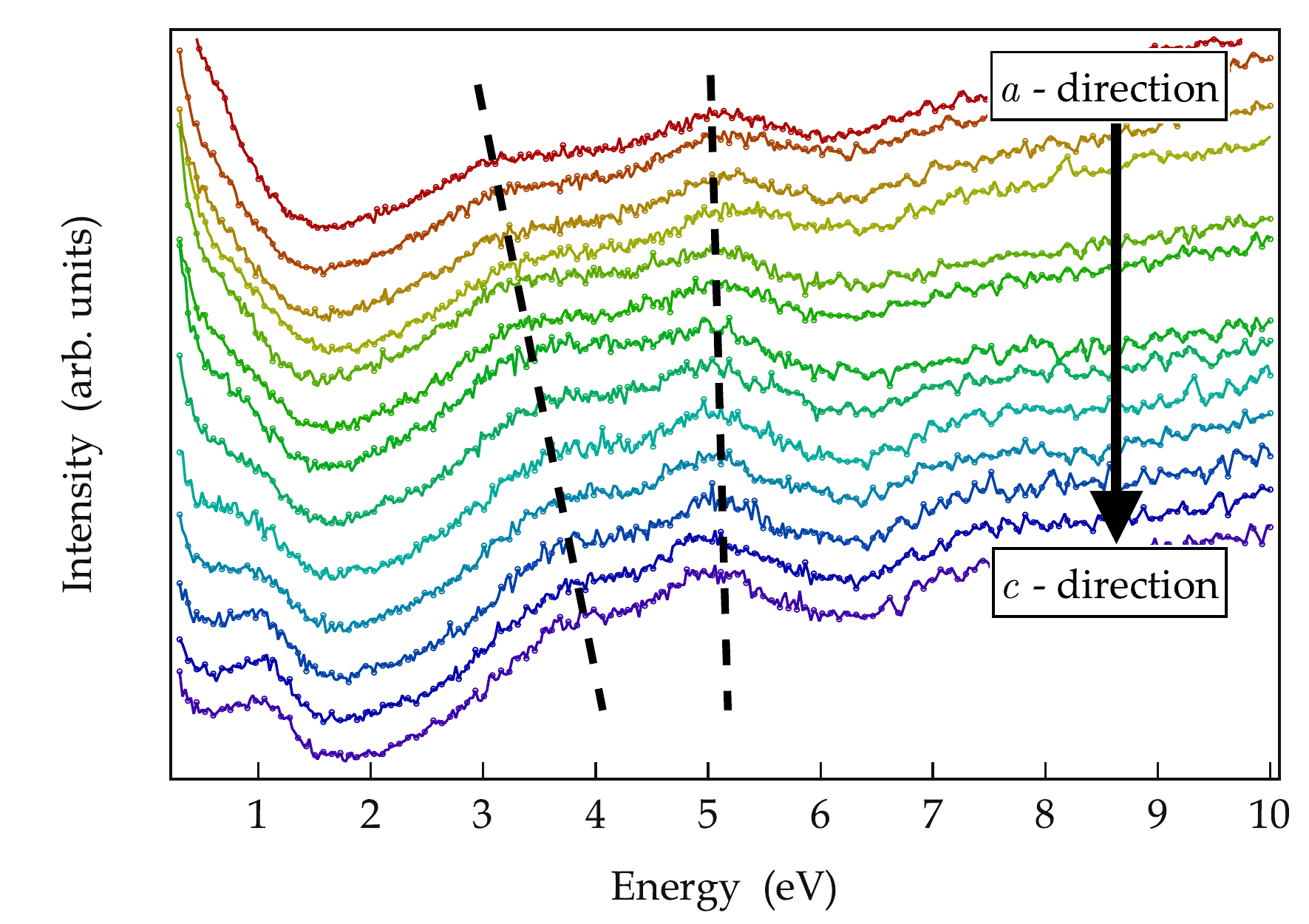}
\caption{\label{fig5}Angular dependence of the EELS response of Ca$_{11}$Sr$_3$Cu$_{24}$O$_{41}$ for $q$ = 0.15\,\AA$^{-1}$ measured in the low
energy range between 0\,-\,10\,eV. The upper spectrum (red line) corresponds to $q \parallel$ $a$, and the lower spectra (purple line)
represents $q \parallel$ $c$.}
\end{figure}

\begin{figure}
\includegraphics[width=0.7\textwidth]{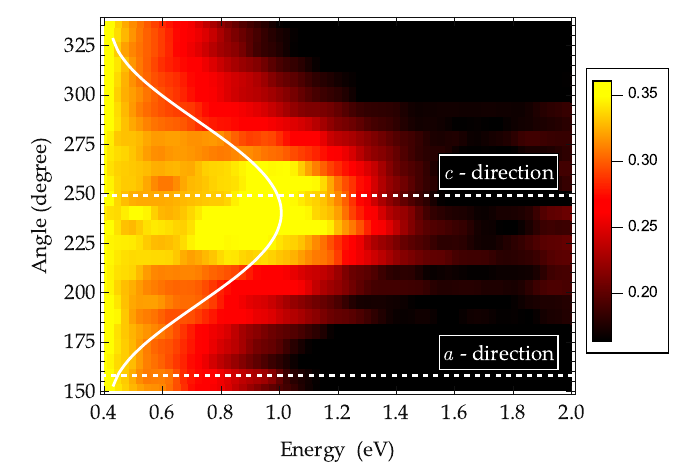}
\caption{\label{fig6}Angular dependence of the EELS intensity for $q$ = 0.15\,\AA$^{-1}$ measured at room temperature in the range between
0.4\,-\,2\,eV. The horizontal dashed bars correspond to the two main crystallographic directions along or perpendicular to the $c$\,-\, and
$a$\,-\, axis, respectively}
\end{figure}

\noindent In order to obtain a more detailed picture we have measured the loss function of Ca$_{11}$Sr$_3$Cu$_{24}$O$_{41}$ as a function of the
angle in the $ac$\,-\,plane at a constant momentum transfer of $q$\,=\,0.15\,\AA$^{-1}$, as presented in Fig.\,\ref{fig5} and \ref{fig6}. We can
identify a clearly visible anisotropy within these directions. In particular, the excitation at 3 - 4\,eV shifts to higher energy approaching
the $c$\,-\,direction, while the spectral feature at 5\,eV remains located at about 5\,eV. The excitation seen at 1\,eV for a momentum transfer
parallel to the $c$\,-\,direction, shows a distinct behavior. Its energy decreases on leaving the $c$-direction and it becomes invisible near
the $a$\,-\,direction. The observed energy variation scales as $\omega_p (\Theta)$\,=\,$\cos (\Theta) \cdot \omega_p$, which follows the
prediction from random-phase-approximation calculations for the charge carrier plasmon excitation of a quasi-one-dimensional
metal.\cite{Williams1974} Resistivity data \cite{Motoyama1997,Kojima2001} also demonstrate a strong anisotropy of the conductivity with a
metallic behavior in $c$\,-\,direction. These facts represent strong support for our conclusion above, that this excitation indeed is a result
of the plasmon oscillation of the
charge carriers in the Cu-O ladder network.\\

\noindent In the following we discuss the origin and character of
the observed excitation in the energy range of 3\,-\,5\,eV. In
this context it is important to consider the structure of
Ca$_{11}$Sr$_3$Cu$_{24}$O$_{41}$. Between two nearest neighbor
copper atoms there are essentially two different
bond-configurations possible in cuprates, which differ in the
angle between the two copper atoms and the relevant $p$-orbital(s)
of a bridging oxygen, a 180$^\circ$ and a 90$^\circ$
bond-configuration. In case of the ladders, the 180$^\circ$
bond-configuration form the legs and rungs of the ladder, while
the copper atoms on neighboring ladders are connected via the
90$^\circ$ bond-configuration (see also Fig.\,\ref{fig1}). The Cu-O
chains in Ca$_{11}$Sr$_3$Cu$_{24}$O$_{41}$ are built from edge
sharing CuO$_4$ units, i.\,e. two copper atoms are connected via
90$^\circ$ Cu-O-Cu bonds. The latter bonding geometry does hardly
allow delocalization of electronic states since hopping of a hole
(or electron) along the chain involves change of the orbital at
the oxygen site. This for instance causes the electronic
excitations to be localized (non-dispersing) as seen for undoped
chains in Li$_2$CuO$_2$.\cite{Atzkern2000} Moreover, the
excitations in these undoped chains also are virtually isotropic
within the plane of the CuO$_4$ units. In the case of doping such
chains the resulting electronic states and the excitations will
still be localized due to the bonding geometry, which is also
evidenced by the results of x-ray absorption studies.\cite{Hu2002}
We therefore attribute the excitation at 5 eV in
Ca$_{11}$Sr$_3$Cu$_{24}$O$_{41}$, which does not disperse and
which is isotropic within the $ac$\,-\,plane, to excitations from
the Cu-O chains in the compound.\\

\noindent In contrast, in the 180$^\circ$ bond-configuration a delocalized charge-transfer excitation is also possible in undoped cuprates. This represents an excited hole which has moved to the O2$p$ states of a neighbouring Cu-O plaquette forming a Zhang-Rice singlet state
there.\cite{Zhang1998,Wang1996,Zhang1988} In addition, this delocalized excitation has a lower excitation energy because of the energy gain
associated with the formation of the Zhang-Rice singlet. In consideration of the structure of the Cu-O ladders, such excitations are most likely
also anisotropic for momentum transfers along and perpendicular to the ladder direction. Although their spectral weight is reduced upon doping, they remain present up to rather high doping levels (about 10\,-\,15\% in planar cuprates \cite{Nuecker1991,Schuster2010}). Thus, the excitation
between 3\,-\,4\,eV is most likely a result from electronic excitations in the Cu-O ladders. The results of recent resonant inelastic x-ray
scattering (RIXS) of Ca$_{x}$Sr$_{14-x}$Cu$_{24}$O$_{41}$ \cite{Wray2008,Wray2008_2,Ishii2007} are in very good agreement with our assignment of
the spectral structures at 3 to 5 eV as seen by EELS. We note that, ignoring the resonance process in RIXS, EELS and RIXS probe the same
dynamic response function. It is therefore very surprising that the available RIXS data on Ca$_{11.5}$Sr$_{2.5}$Cu$_{24}$O$_{41}$ miss the low
energy plasmon excitation at 1\,eV. It is unclear at present whether this is due
to limitations in the experimental parameters such as resolution, or whether it has an intrinsic origin, related to the resonant process itself.\\

\noindent Finally, the dispersion of the charge carrier plasmon
can also help to gain further insight into the microscopic nature
of the electronic system. For a simple metal in the long
wavelength limit, the plasmon dispersion is expected to be
quadratic, i.\,e. $\omega (q)\,=\,\omega_p + \mathrm{A}q^2$, whereas the
coefficient A can be expressed as $\mathrm{A}\,=\,\frac{\hbar^2}{m}\alpha$ and $\alpha = \alpha_1 +
\alpha_2$ \cite{Sing1999,SingPhD}, with

\begin{align}
\alpha_1 = \frac{m v_F^2}{2 \hbar \omega_p} \qquad \mathrm{and} \qquad \alpha_2 = \frac{me^2}{24 \hbar\epsilon_{\infty}\epsilon_0 \omega_p}
\left\langle v_q \left(\textbf{e} \frac{\partial}{\partial\textbf{k}}\right)^2 v_q \right\rangle.
\end{align}

\noindent Thereby $m$ is the free-electron mass, $v_F$ is the Fermi velocity, $\epsilon_{\infty}$ is the background dielectric constant and
$\omega_p$ is the plasma frequency. To calculate $\alpha_2$ one has to consider the second derivative of the Fermi velocity component parallel
to $q$, $v_q$.

\par

Interestingly, the dispersion of the plasmon in
Ca$_{11}$Sr$_3$Cu$_{24}$O$_{41}$ along the ladder direction (see
Fig.\,\ref{fig4}) has a band width which is comparable to that
observed for the optimally doped high temperature superconductors.
This seems reasonable, taking into account that the doping level of
the ladders in Ca$_{11}$Sr$_3$Cu$_{24}$O$_{41}$ is about 0.15 -
0.2 holes per Cu unit, as reported from recent angular resolved
photoemission experiments \cite{Koitzsch2010}. However, the
plasmon in Ca$_{11}$Sr$_3$Cu$_{24}$O$_{41}$ scales linearly with
$q$, which is in contrast to the high T$_c$ materials at optimal
doping, and also in contrast to the expectation for a free
electron gas like electronic system.

\par

In order to investigate the long wave length limit of the plasmon
dispersion in Ca$_{11}$Sr$_3$Cu$_{24}$O$_{41}$ in more detail, we
have fitted the dispersion curve using a polynomial function

\begin{align}
\omega(q)\,=\,\omega_p + \mathrm{A}q^2 + \mathrm{B}q^4.
\end{align}

The parameter A then represents the plasmon behavior for small momenta, i.\,e. long wavelengths. This fit provides us with the following results
(see also Fig.\,\ref{fig4}): $\hbar\omega_p$ = (0.83$\pm$0.02)\,eV, A\,=\,(7.47$\pm$0.75)\,eV\AA$^2$ and
B\,=\,(-25.28$\pm$5.48)\,eV\AA$^4$.

 For a one-dimensional system the plasma frequency $\omega_p$ can be written as \cite{Sing1999}

\begin{align}
\omega_p^2 = \frac{4e^2}{\hbar \epsilon_{\infty} \epsilon_0 \pi a b} |v_F|.
\end{align}

Note that this expression takes the number of Cu-sites within the
unit cell of the ladder into account.

\par

\begin{figure}[ht]
 \includegraphics[width=0.45\textwidth]{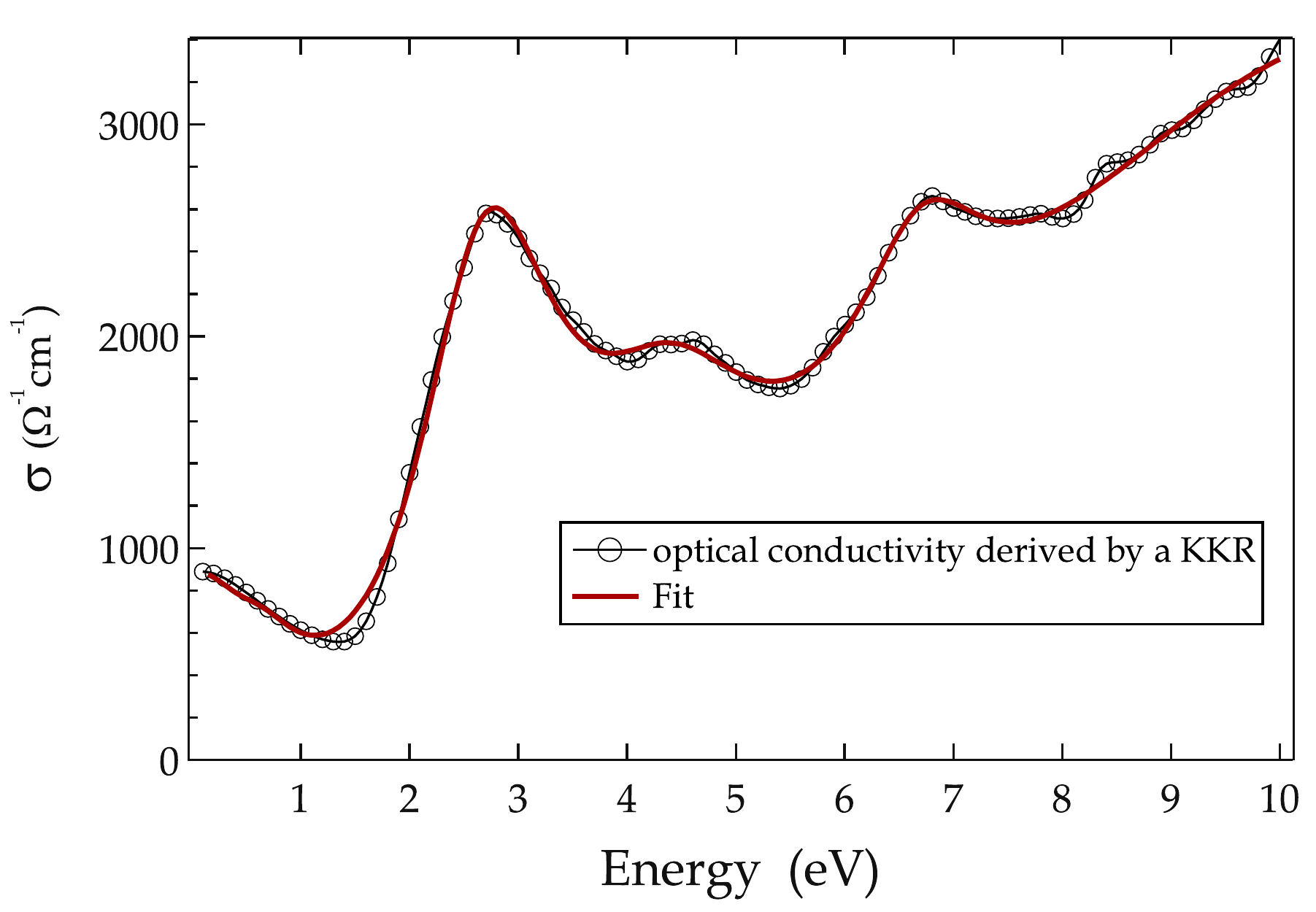}
 \includegraphics[width=0.45\textwidth]{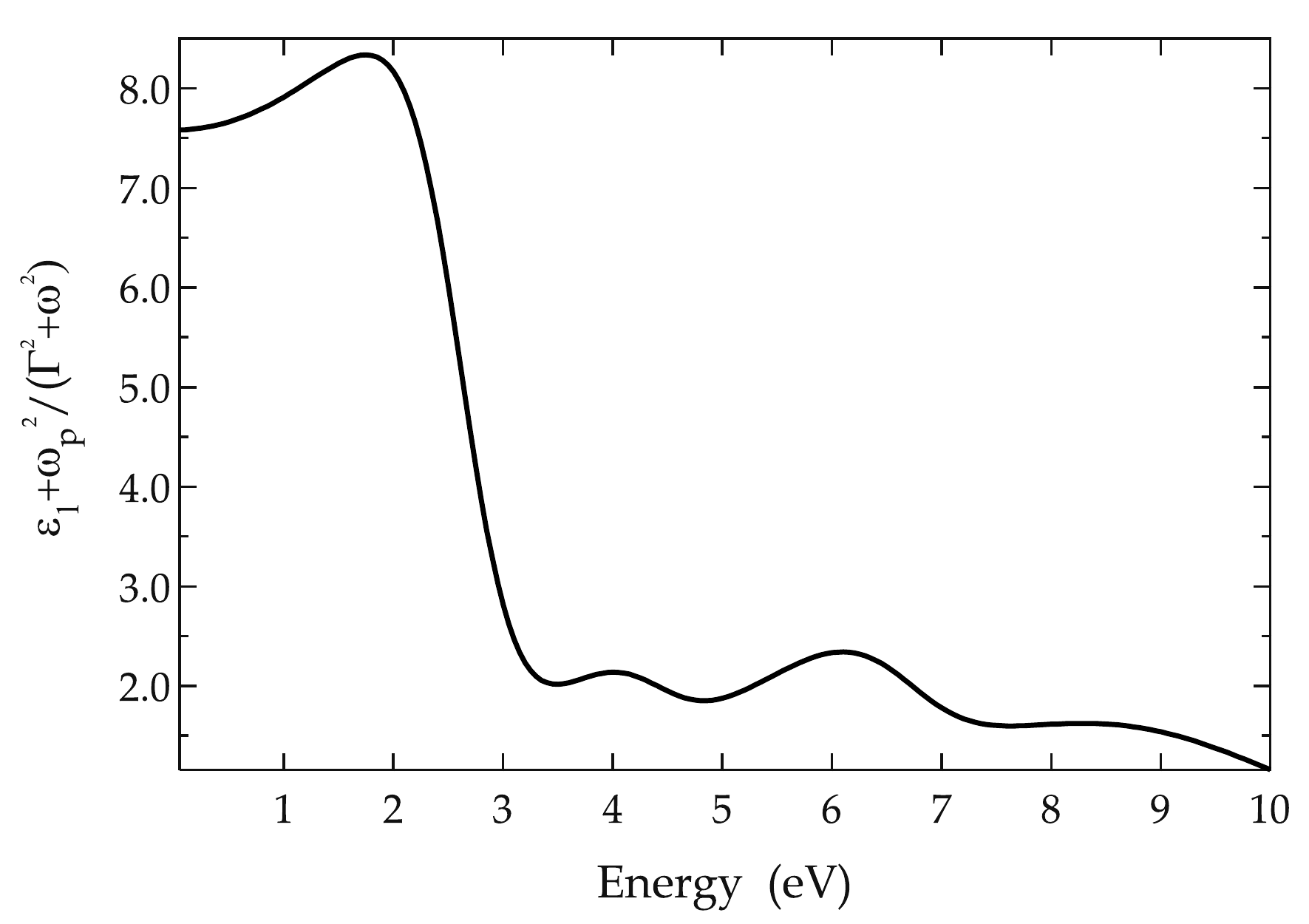}
\caption{\label{fig7}Left panel: The optical conductivity for Ca$_{11}$Sr$_{3}$Cu$_{24}$O$_{41}$ as derived by a Kramers-Kronig transformation of the EELS intensity (black circle) and the fit (red line). Right panel: The real part of the dielectric function  subtracted by the Drude contribution which provides a value for the background dielectric constant of $\epsilon_{\infty}$\,$\approx$\,7.6.}
\end{figure}

To be able to derive the mean Fermi velocity from the expression above, the knowledge of the background dielectric constant $\epsilon_{\infty}$
is required. We have analyzed this parameter via a Kramers-Kronig analysis (KKA) of the measured loss function. Subsequently, we have described
the resulting dielectric function within the Drude-Lorentz model with one Drude and a number of Lorentz oscillators. In left panel of
Fig.\,\ref{fig7} we show the optical conductivity ($\sigma = \omega \epsilon_2$) as received from our KKA and the corresponding fit result. This
Figure demonstrates that our model description of the data is very good. The value of the plasma frequency $\omega_p$\,=\,0.89\,eV which is
obtained from the fit of $\sigma$ is in very good agreement to that provided by the fit of the plasmon dispersion (cf. Fig.\,\ref{fig4}). In addition, this value is also in good agreement to the data from reflectivity measurements.\cite{Ruzicka1998}

The background dielectric constant can now be read off the real part of the dielectric function subtracting the Drude (i.\,e. charge carrier)
contribution (see Fig.\,\ref{fig7} right panel), we obtain $\epsilon_{\infty}$\,$\approx$\,7.5 - 8. We thus arrive at a mean Fermi velocity for the conduction electrons in the ladder of
$v_F\,\approx$\,530000\,$\frac{\mathrm{m}}{\mathrm{s}}$. Taking this value, we can now calculate the parameter $\alpha_1$ to be about
$\sim$\,0.96 (and consequently a value for A of 7.25\,eV\AA$^2$), which is very close to what we obtained from the fit for this coefficient,
A\,=\,7.47\,eV\AA$^2$.

This good correspondence indicates that (i) our description of the long wavelenght limit is consistent and (ii) the contribution of $\alpha_2$
to the long wave length plasmon dispersion is small. Indeed, a calculation of $\alpha_2$ using equation (2) and taking into account the tight
binding description of the conduction bands in Ca$_{11}$Sr$_3$Cu$_{24}$O$_{41}$ \cite{Arai1997} yields \(|\alpha_2|\,\le\,0.05\) (for a more
detailed description of this procedure see \cite{Sing1999_2}). Thus, $\alpha_2$ is less than 10$\%$ of the contribution to the dispersion
coefficient A.

We now can conclude that the long wavelength limit of the plasmon
dispersion in Ca$_{11}$Sr$_3$Cu$_{24}$O$_{41}$ can be well
rationalized within an RPA like description of simple metals with
a mean Fermi velocity of 530000\,$\frac{\mathrm{m}}{\mathrm{s}}$, which is in reasonable agreement
with the tight binding description of the conduction bands in
Ca$_{11}$Sr$_3$Cu$_{24}$O$_{41}$ \cite{Arai1997}, and which also agrees well with
recent result from angular resolved photoemission.\cite{Koitzsch2010}

\par

However, the quasi-linear dispersion as revealed in Fig. \ref{fig4} cannot
be rationalized within the framework of a simple metal. Deviations
from the expectation of a quadratic plasmon dispersion have
already been reported in the past. Already the heavier alkali
metals show vanishing or even negative plasmon dispersions, which
has initiated a lot of theoretical work.\cite{Felde1989} Over the years different
reasons for these observations have been discussed including local
field effects, interband transitions and the anisotropy of the
effective mass. This emphasizes that the dispersion of the charge
carrier plasmon can be a very complex parameter.

\par

Previously, the plasmon dispersion in other quasi one-dimensional metallic systems has been investigated theoretically and experimentally for a
few compounds. Within RPA it has been predicted \cite{Williams1974} that the plasmon dispersion in one dimensional metals can be substantially
modified by local field effects, i.\,e. the inhomogeneous character of the electron gas. This modification can even cause a negative plasmon
dispersion in case of a tight binding description of the electronic bands.\cite{Williams1974} Experimental studies of the plasmon dispersion in
(TaSe$_4$)$_2$I \cite{Sing1998} and K$_{0.3}$MoO$_3$, \cite{Sing1999} found a quasi-linear dispersion which could be explained to predominantly
be an effect of the band structure in these materials.

Moreover, going to lower doping levels of about 0.1 holes per Cu atom in two-dimensional cuprate structures, the plasmon dispersion is  drastically reduced compared to optimal doping (i.\,e. about 0.15 holes per copper atom). The band width of the plasmon in
Ca$_{1.9}$Na$_{0.1}$CuO$_2$Cl$_2$ is only half of that observed for the doping level of 0.15 holes per copper unit.\cite{Schuster2010} Also, the
plasmon dispersion at this lower doping in planar cuprates is essentially linear in contrast to optimal doping, but reminiscent of our results
for Ca$_{11}$Sr$_3$Cu$_{24}$O$_{41}$. Since at lower doping (so-called underdoping) the planar cuprates enter a pseudogap phase, the origin of
which is actually under debate, the variations of the plasmon behavior might be closely connected with the peculiar properties in this
pseudogap region. In this context, it is important to notice that there is evidence that the electronic degrees of freedom in the cuprate
ladders are also quite complex. The data from angular resolved photoemission \cite{Koitzsch2010} show a substantially reduced spectral weight
close to the Fermi level, and the optical reflectivity \cite{Osafune1997, Ruzicka1998} is different from that of a simple metallic material, but
indicates additional electronic excitation in the energy range of the plasmon. In addition, for Ca$_{x}$Sr$_{14-x}$Cu$_{24}$O$_{41}$ compounds
the formation of a hole crystal \cite{Abbamonte2004,Carr2002,Friedel2002,Hess2004} (i.\,e. a charge density wave) has been reported. These findings
suggest that also in the cuprate ladders there might be a phase quite similar to the pseudo gap phase in the planar cuprates, with complex
electronic degrees of freedom and interactions.

\par

\noindent In relation to these previous findings we conclude that in
Ca$_{11}$Sr$_3$Cu$_{24}$O$_{41}$, band structure and local field
effects as well as the peculiar physics of underdoped cuprates
have to be considered in the future, in order to rationalize the
measured plasmon dispersion, and further theoretical developments
are required to achieve a conclusive picture of this interesting
physics.

\section{Summary}
\noindent To summarize, employing EELS we investigated the dispersion of low lying charge-transfer excitations and the charge carrier plasmon in
the spin ladder system Ca$_{11}$Sr$_3$Cu$_{24}$O$_{41}$. We found a strong anisotropy of the spectral structures, whereas the charge carrier
plasmon is only visible for a momentum parallel to the crystallographic $c$ - direction. A well pronounced two peak structure is seen at  3\,-\,5\,eV, and can be qualitatively assigned to localized and delocalized charge-transfer excitations. The plasmon dispersion scales quasi linear along the
legs of the ladder, which is in contrast to what is observed for cuprate high temperature superconductors. A comparison of the fit of the
dispersion curve (using a polynomial function) with the value for the plasma frequency (which can be obtained by a fit of the optical conductivity
after a Kramers Kronig analysis of the measured loss function) shows a very good agreement and consistency of both fits. This indicates that
the long wavelenght limit of the plasmon dispersion can be described within a RPA like description. We furthermore calculated the mean Fermi
velocity to be about 530000\,$\frac{\mathrm{m}}{\mathrm{s}}$, which agrees well with a tight binding description of
Ca$_{11}$Sr$_3$Cu$_{24}$O$_{41}$ and with results from angular resolved photoemission. The linearity of the plasmon dispersion cannot be
rationalized within the framework of a simple metal. Phenomena such as local fields, interband transistions, or the influence of the band
structure, as well as many-body effects in cuprates, may be responsible for this behavior.

\begin{acknowledgments}
\noindent We thank R. Sch\"onfelder, R. H\"ubel and S. Leger for technical
assistance. This work has been supported by the Deutsche
Forschungsgemeinschaft (grant number KN 393/13).
\end{acknowledgments}

\bibliography{bib_ladder}

\begin{thebibliography}{54}%
\makeatletter
\providecommand \@ifxundefined [1]{%
 \@ifx{#1\undefined}
}%
\providecommand \@ifnum [1]{%
 \ifnum #1\expandafter \@firstoftwo
 \else \expandafter \@secondoftwo
 \fi
}%
\providecommand \@ifx [1]{%
 \ifx #1\expandafter \@firstoftwo
 \else \expandafter \@secondoftwo
 \fi
}%
\providecommand \natexlab [1]{#1}%
\providecommand \enquote  [1]{``#1''}%
\providecommand \bibnamefont  [1]{#1}%
\providecommand \bibfnamefont [1]{#1}%
\providecommand \citenamefont [1]{#1}%
\providecommand \href@noop [0]{\@secondoftwo}%
\providecommand \href [0]{\begingroup \@sanitize@url \@href}%
\providecommand \@href[1]{\@@startlink{#1}\@@href}%
\providecommand \@@href[1]{\endgroup#1\@@endlink}%
\providecommand \@sanitize@url [0]{\catcode `\\12\catcode `\$12\catcode
  `\&12\catcode `\#12\catcode `\^12\catcode `\_12\catcode `\%12\relax}%
\providecommand \@@startlink[1]{}%
\providecommand \@@endlink[0]{}%
\providecommand \url  [0]{\begingroup\@sanitize@url \@url }%
\providecommand \@url [1]{\endgroup\@href {#1}{\urlprefix }}%
\providecommand \urlprefix  [0]{URL }%
\providecommand \Eprint [0]{\href }%
\@ifxundefined \urlstyle {%
  \providecommand \doi  [0]{\begingroup \@sanitize@url \@doi}%
  \providecommand \@doi [1]{\endgroup \@@startlink {\doibase
  #1}doi:\discretionary {}{}{}#1\@@endlink }%
}{%
  \providecommand \doi  [0]{doi:\discretionary{}{}{}\begingroup
  \urlstyle{rm}\Url }%
}%
\providecommand \doibase [0]{http://dx.doi.org/}%
\providecommand \Doi [0]{\begingroup \@sanitize@url \@Doi }%
\providecommand \@Doi  [1]{\endgroup\@@startlink{\doibase#1}\@@Doi}%
\providecommand \@@Doi [1]{#1\@@endlink}%
\providecommand \selectlanguage [0]{\@gobble}%
\providecommand \bibinfo  [0]{\@secondoftwo}%
\providecommand \bibfield  [0]{\@secondoftwo}%
\providecommand \translation [1]{[#1]}%
\providecommand \BibitemOpen [0]{}%
\providecommand \bibitemStop [0]{}%
\providecommand \bibitemNoStop [0]{.\EOS\space}%
\providecommand \EOS [0]{\spacefactor3000\relax}%
\providecommand \BibitemShut  [1]{\csname bibitem#1\endcsname}%
\bibitem [{\citenamefont {Bednorz}\ and\ \citenamefont
  {M{\"u}ller}(1986)}]{Bednorz1986}%
  \BibitemOpen
  \bibfield  {author} {\bibinfo {author} {\bibfnamefont {J.}~\bibnamefont
  {Bednorz}}\ and\ \bibinfo {author} {\bibfnamefont {K.}~\bibnamefont
  {M{\"u}ller}},\ }\href@noop {} {\bibfield  {journal} {\bibinfo  {journal} {Z.
  Phys. B},\ }\textbf {\bibinfo {volume} {64}},\ \bibinfo {pages} {189}
  (\bibinfo {year} {1986})}\BibitemShut {NoStop}%
\bibitem [{\citenamefont {McCarron}\ \emph {et~al.}(1988)\citenamefont
  {McCarron}, \citenamefont {Subramanian}, \citenamefont {Calabrese},\ and\
  \citenamefont {Harlow}}]{McCarron1998}%
  \BibitemOpen
  \bibfield  {author} {\bibinfo {author} {\bibfnamefont {E.}~\bibnamefont
  {McCarron}}, \bibinfo {author} {\bibfnamefont {M.}~\bibnamefont
  {Subramanian}}, \bibinfo {author} {\bibfnamefont {J.}~\bibnamefont
  {Calabrese}}, \ and\ \bibinfo {author} {\bibfnamefont {R.}~\bibnamefont
  {Harlow}},\ }\href@noop {} {\bibfield  {journal} {\bibinfo  {journal} {Mater.
  Res. Bull.},\ }\textbf {\bibinfo {volume} {23}},\ \bibinfo {pages} {1355}
  (\bibinfo {year} {1988})}\BibitemShut {NoStop}%
\bibitem [{\citenamefont {Siegrist}\ \emph {et~al.}(1988)\citenamefont
  {Siegrist}, \citenamefont {Schneemeyer}, \citenamefont {Sunshine},
  \citenamefont {Waszczak},\ and\ \citenamefont {Roth}}]{Siegrist1988}%
  \BibitemOpen
  \bibfield  {author} {\bibinfo {author} {\bibfnamefont {T.}~\bibnamefont
  {Siegrist}}, \bibinfo {author} {\bibfnamefont {L.}~\bibnamefont
  {Schneemeyer}}, \bibinfo {author} {\bibfnamefont {S.}~\bibnamefont
  {Sunshine}}, \bibinfo {author} {\bibfnamefont {J.}~\bibnamefont {Waszczak}},
  \ and\ \bibinfo {author} {\bibfnamefont {R.}~\bibnamefont {Roth}},\
  }\href@noop {} {\bibfield  {journal} {\bibinfo  {journal} {Mater. Res.
  Bull.},\ }\textbf {\bibinfo {volume} {23}},\ \bibinfo {pages} {1429}
  (\bibinfo {year} {1988})}\BibitemShut {NoStop}%
\bibitem [{\citenamefont {Uehara}\ \emph {et~al.}(1996)\citenamefont {Uehara},
  \citenamefont {Nagata}, \citenamefont {Akimitsu}, \citenamefont {Takahashi},
  \citenamefont {Mori},\ and\ \citenamefont {Kinoshita}}]{Uehara1996}%
  \BibitemOpen
  \bibfield  {author} {\bibinfo {author} {\bibfnamefont {M.}~\bibnamefont
  {Uehara}}, \bibinfo {author} {\bibfnamefont {T.}~\bibnamefont {Nagata}},
  \bibinfo {author} {\bibfnamefont {J.}~\bibnamefont {Akimitsu}}, \bibinfo
  {author} {\bibfnamefont {H.}~\bibnamefont {Takahashi}}, \bibinfo {author}
  {\bibfnamefont {N.}~\bibnamefont {Mori}}, \ and\ \bibinfo {author}
  {\bibfnamefont {K.}~\bibnamefont {Kinoshita}},\ }\href@noop {} {\bibfield
  {journal} {\bibinfo  {journal} {J. Phys. Soc. Jpn.},\ }\textbf {\bibinfo
  {volume} {65}},\ \bibinfo {pages} {2764} (\bibinfo {year}
  {1996})}\BibitemShut {NoStop}%
\bibitem [{\citenamefont {Nagata}\ \emph {et~al.}(1997)\citenamefont {Nagata},
  \citenamefont {Uehara}, \citenamefont {Goto}, \citenamefont {Komiya},
  \citenamefont {Akimitsu}, \citenamefont {Motoyama}, \citenamefont {Eisaki},
  \citenamefont {Uchida}, \citenamefont {Takahashi}, \citenamefont
  {Nakanishi},\ and\ \citenamefont {Môri}}]{Nagata1997}%
  \BibitemOpen
  \bibfield  {author} {\bibinfo {author} {\bibfnamefont {T.}~\bibnamefont
  {Nagata}}, \bibinfo {author} {\bibfnamefont {M.}~\bibnamefont {Uehara}},
  \bibinfo {author} {\bibfnamefont {J.}~\bibnamefont {Goto}}, \bibinfo {author}
  {\bibfnamefont {N.}~\bibnamefont {Komiya}}, \bibinfo {author} {\bibfnamefont
  {J.}~\bibnamefont {Akimitsu}}, \bibinfo {author} {\bibfnamefont
  {N.}~\bibnamefont {Motoyama}}, \bibinfo {author} {\bibfnamefont
  {H.}~\bibnamefont {Eisaki}}, \bibinfo {author} {\bibfnamefont
  {S.}~\bibnamefont {Uchida}}, \bibinfo {author} {\bibfnamefont
  {H.}~\bibnamefont {Takahashi}}, \bibinfo {author} {\bibfnamefont
  {T.}~\bibnamefont {Nakanishi}}, \ and\ \bibinfo {author} {\bibfnamefont
  {N.}~\bibnamefont {Môri}},\ }\Doi {DOI: 10.1016/S0921-4534(97)00247-5}
  {\bibfield  {journal} {\bibinfo  {journal} {Physica C: Superconductivity},\
  }\textbf {\bibinfo {volume} {282-287}},\ \bibinfo {pages} {153 } (\bibinfo
  {year} {1997})}\BibitemShut {NoStop}%
\bibitem [{\citenamefont {Kato}\ \emph {et~al.}(1996)\citenamefont {Kato},
  \citenamefont {Shiota},\ and\ \citenamefont {Koike}}]{Kato1996}%
  \BibitemOpen
  \bibfield  {author} {\bibinfo {author} {\bibfnamefont {M.}~\bibnamefont
  {Kato}}, \bibinfo {author} {\bibfnamefont {K.}~\bibnamefont {Shiota}}, \ and\
  \bibinfo {author} {\bibfnamefont {Y.}~\bibnamefont {Koike}},\ }\Doi {DOI:
  10.1016/0921-4534(95)00802-0} {\bibfield  {journal} {\bibinfo  {journal}
  {Physica C: Superconductivity},\ }\textbf {\bibinfo {volume} {258}},\
  \bibinfo {pages} {284 } (\bibinfo {year} {1996})}\BibitemShut {NoStop}%
\bibitem [{\citenamefont {N\"ucker}\ \emph {et~al.}(2000)\citenamefont
  {N\"ucker}, \citenamefont {Merz}, \citenamefont {Kuntscher}, \citenamefont
  {Gerhold}, \citenamefont {Schuppler}, \citenamefont {Neudert}, \citenamefont
  {Golden}, \citenamefont {Fink}, \citenamefont {Schild}, \citenamefont
  {Stadler}, \citenamefont {Chakarian}, \citenamefont {Freeland}, \citenamefont
  {Idzerda}, \citenamefont {Conder}, \citenamefont {Uehara}, \citenamefont
  {Nagata}, \citenamefont {Goto}, \citenamefont {Akimitsu}, \citenamefont
  {Motoyama}, \citenamefont {Eisaki}, \citenamefont {Uchida}, \citenamefont
  {Ammerahl},\ and\ \citenamefont {Revcolevschi}}]{Nuecker2000}%
  \BibitemOpen
  \bibfield  {author} {\bibinfo {author} {\bibfnamefont {N.}~\bibnamefont
  {N\"ucker}}, \bibinfo {author} {\bibfnamefont {M.}~\bibnamefont {Merz}},
  \bibinfo {author} {\bibfnamefont {C.~A.}\ \bibnamefont {Kuntscher}}, \bibinfo
  {author} {\bibfnamefont {S.}~\bibnamefont {Gerhold}}, \bibinfo {author}
  {\bibfnamefont {S.}~\bibnamefont {Schuppler}}, \bibinfo {author}
  {\bibfnamefont {R.}~\bibnamefont {Neudert}}, \bibinfo {author} {\bibfnamefont
  {M.~S.}\ \bibnamefont {Golden}}, \bibinfo {author} {\bibfnamefont
  {J.}~\bibnamefont {Fink}}, \bibinfo {author} {\bibfnamefont {D.}~\bibnamefont
  {Schild}}, \bibinfo {author} {\bibfnamefont {S.}~\bibnamefont {Stadler}},
  \bibinfo {author} {\bibfnamefont {V.}~\bibnamefont {Chakarian}}, \bibinfo
  {author} {\bibfnamefont {J.}~\bibnamefont {Freeland}}, \bibinfo {author}
  {\bibfnamefont {Y.~U.}\ \bibnamefont {Idzerda}}, \bibinfo {author}
  {\bibfnamefont {K.}~\bibnamefont {Conder}}, \bibinfo {author} {\bibfnamefont
  {M.}~\bibnamefont {Uehara}}, \bibinfo {author} {\bibfnamefont
  {T.}~\bibnamefont {Nagata}}, \bibinfo {author} {\bibfnamefont
  {J.}~\bibnamefont {Goto}}, \bibinfo {author} {\bibfnamefont {J.}~\bibnamefont
  {Akimitsu}}, \bibinfo {author} {\bibfnamefont {N.}~\bibnamefont {Motoyama}},
  \bibinfo {author} {\bibfnamefont {H.}~\bibnamefont {Eisaki}}, \bibinfo
  {author} {\bibfnamefont {S.}~\bibnamefont {Uchida}}, \bibinfo {author}
  {\bibfnamefont {U.}~\bibnamefont {Ammerahl}}, \ and\ \bibinfo {author}
  {\bibfnamefont {A.}~\bibnamefont {Revcolevschi}},\ }\Doi
  {10.1103/PhysRevB.62.14384} {\bibfield  {journal} {\bibinfo  {journal} {Phys.
  Rev. B},\ }\textbf {\bibinfo {volume} {62}},\ \bibinfo {pages} {14384}
  (\bibinfo {year} {2000})}\BibitemShut {NoStop}%
\bibitem [{\citenamefont {Koitzsch}\ \emph {et~al.}(2010)\citenamefont
  {Koitzsch}, \citenamefont {Inosov}, \citenamefont {Shiozawa}, \citenamefont
  {Zabolotnyy}, \citenamefont {Borisenko}, \citenamefont {Varykhalov},
  \citenamefont {Hess}, \citenamefont {Knupfer}, \citenamefont {Ammerahl},
  \citenamefont {Revcolevschi},\ and\ \citenamefont
  {B\"uchner}}]{Koitzsch2010}%
  \BibitemOpen
  \bibfield  {author} {\bibinfo {author} {\bibfnamefont {A.}~\bibnamefont
  {Koitzsch}}, \bibinfo {author} {\bibfnamefont {D.~S.}\ \bibnamefont
  {Inosov}}, \bibinfo {author} {\bibfnamefont {H.}~\bibnamefont {Shiozawa}},
  \bibinfo {author} {\bibfnamefont {V.~B.}\ \bibnamefont {Zabolotnyy}},
  \bibinfo {author} {\bibfnamefont {S.~V.}\ \bibnamefont {Borisenko}}, \bibinfo
  {author} {\bibfnamefont {A.}~\bibnamefont {Varykhalov}}, \bibinfo {author}
  {\bibfnamefont {C.}~\bibnamefont {Hess}}, \bibinfo {author} {\bibfnamefont
  {M.}~\bibnamefont {Knupfer}}, \bibinfo {author} {\bibfnamefont
  {U.}~\bibnamefont {Ammerahl}}, \bibinfo {author} {\bibfnamefont
  {A.}~\bibnamefont {Revcolevschi}}, \ and\ \bibinfo {author} {\bibfnamefont
  {B.}~\bibnamefont {B\"uchner}},\ }\href@noop {} {\bibfield  {journal}
  {\bibinfo  {journal} {Phys. Rev. B},\ }\textbf {\bibinfo {volume} {81}},\
  \bibinfo {pages} {113110} (\bibinfo {year} {2010})}\BibitemShut {NoStop}%
\bibitem [{\citenamefont {Kataev}\ \emph {et~al.}(2001)\citenamefont {Kataev},
  \citenamefont {Choi}, \citenamefont {Gr\"uninger}, \citenamefont {Ammerahl},
  \citenamefont {B\"uchner}, \citenamefont {Freimuth},\ and\ \citenamefont
  {Revcolevschi}}]{Kataev2001}%
  \BibitemOpen
  \bibfield  {author} {\bibinfo {author} {\bibfnamefont {V.}~\bibnamefont
  {Kataev}}, \bibinfo {author} {\bibfnamefont {K.-Y.}\ \bibnamefont {Choi}},
  \bibinfo {author} {\bibfnamefont {M.}~\bibnamefont {Gr\"uninger}}, \bibinfo
  {author} {\bibfnamefont {U.}~\bibnamefont {Ammerahl}}, \bibinfo {author}
  {\bibfnamefont {B.}~\bibnamefont {B\"uchner}}, \bibinfo {author}
  {\bibfnamefont {A.}~\bibnamefont {Freimuth}}, \ and\ \bibinfo {author}
  {\bibfnamefont {A.}~\bibnamefont {Revcolevschi}},\ }\Doi
  {10.1103/PhysRevLett.86.2882} {\bibfield  {journal} {\bibinfo  {journal}
  {Phys. Rev. Lett.},\ }\textbf {\bibinfo {volume} {86}},\ \bibinfo {pages}
  {2882} (\bibinfo {year} {2001})}\BibitemShut {NoStop}%
\bibitem [{\citenamefont {Ammerahl}\ \emph {et~al.}(2000)\citenamefont
  {Ammerahl}, \citenamefont {B\"uchner}, \citenamefont {Colonescu},
  \citenamefont {Gross},\ and\ \citenamefont {Revcolevschi}}]{Ammerahl2000}%
  \BibitemOpen
  \bibfield  {author} {\bibinfo {author} {\bibfnamefont {U.}~\bibnamefont
  {Ammerahl}}, \bibinfo {author} {\bibfnamefont {B.}~\bibnamefont {B\"uchner}},
  \bibinfo {author} {\bibfnamefont {L.}~\bibnamefont {Colonescu}}, \bibinfo
  {author} {\bibfnamefont {R.}~\bibnamefont {Gross}}, \ and\ \bibinfo {author}
  {\bibfnamefont {A.}~\bibnamefont {Revcolevschi}},\ }\Doi
  {10.1103/PhysRevB.62.8630} {\bibfield  {journal} {\bibinfo  {journal} {Phys.
  Rev. B},\ }\textbf {\bibinfo {volume} {62}},\ \bibinfo {pages} {8630}
  (\bibinfo {year} {2000})}\BibitemShut {NoStop}%
\bibitem [{\citenamefont {Osafune}\ \emph {et~al.}(1997)\citenamefont
  {Osafune}, \citenamefont {Motoyama}, \citenamefont {Eisaki},\ and\
  \citenamefont {Uchida}}]{Osafune1997}%
  \BibitemOpen
  \bibfield  {author} {\bibinfo {author} {\bibfnamefont {T.}~\bibnamefont
  {Osafune}}, \bibinfo {author} {\bibfnamefont {N.}~\bibnamefont {Motoyama}},
  \bibinfo {author} {\bibfnamefont {H.}~\bibnamefont {Eisaki}}, \ and\ \bibinfo
  {author} {\bibfnamefont {S.}~\bibnamefont {Uchida}},\ }\Doi
  {10.1103/PhysRevLett.78.1980} {\bibfield  {journal} {\bibinfo  {journal}
  {Phys. Rev. Lett.},\ }\textbf {\bibinfo {volume} {78}},\ \bibinfo {pages}
  {1980} (\bibinfo {year} {1997})}\BibitemShut {NoStop}%
\bibitem [{\citenamefont {Rusydi}\ \emph {et~al.}(2007)\citenamefont {Rusydi},
  \citenamefont {Berciu}, \citenamefont {Abbamonte}, \citenamefont {Smadici},
  \citenamefont {Eisaki}, \citenamefont {Fujimaki}, \citenamefont {Uchida},
  \citenamefont {R\"ubhausen},\ and\ \citenamefont {Sawatzky}}]{Rusydi2007}%
  \BibitemOpen
  \bibfield  {author} {\bibinfo {author} {\bibfnamefont {A.}~\bibnamefont
  {Rusydi}}, \bibinfo {author} {\bibfnamefont {M.}~\bibnamefont {Berciu}},
  \bibinfo {author} {\bibfnamefont {P.}~\bibnamefont {Abbamonte}}, \bibinfo
  {author} {\bibfnamefont {S.}~\bibnamefont {Smadici}}, \bibinfo {author}
  {\bibfnamefont {H.}~\bibnamefont {Eisaki}}, \bibinfo {author} {\bibfnamefont
  {Y.}~\bibnamefont {Fujimaki}}, \bibinfo {author} {\bibfnamefont
  {S.}~\bibnamefont {Uchida}}, \bibinfo {author} {\bibfnamefont
  {M.}~\bibnamefont {R\"ubhausen}}, \ and\ \bibinfo {author} {\bibfnamefont
  {G.~A.}\ \bibnamefont {Sawatzky}},\ }\Doi {10.1103/PhysRevB.75.104510}
  {\bibfield  {journal} {\bibinfo  {journal} {Phys. Rev. B},\ }\textbf
  {\bibinfo {volume} {75}},\ \bibinfo {pages} {104510} (\bibinfo {year}
  {2007})}\BibitemShut {NoStop}%
\bibitem [{\citenamefont {Magishi}\ \emph {et~al.}(1998)\citenamefont
  {Magishi}, \citenamefont {Matsumoto}, \citenamefont {Kitaoka}, \citenamefont
  {Ishida}, \citenamefont {Asayama}, \citenamefont {Uehara}, \citenamefont
  {Nagata},\ and\ \citenamefont {Akimitsu}}]{Magishi1998}%
  \BibitemOpen
  \bibfield  {author} {\bibinfo {author} {\bibfnamefont {K.}~\bibnamefont
  {Magishi}}, \bibinfo {author} {\bibfnamefont {S.}~\bibnamefont {Matsumoto}},
  \bibinfo {author} {\bibfnamefont {Y.}~\bibnamefont {Kitaoka}}, \bibinfo
  {author} {\bibfnamefont {K.}~\bibnamefont {Ishida}}, \bibinfo {author}
  {\bibfnamefont {K.}~\bibnamefont {Asayama}}, \bibinfo {author} {\bibfnamefont
  {M.}~\bibnamefont {Uehara}}, \bibinfo {author} {\bibfnamefont
  {T.}~\bibnamefont {Nagata}}, \ and\ \bibinfo {author} {\bibfnamefont
  {J.}~\bibnamefont {Akimitsu}},\ }\Doi {10.1103/PhysRevB.57.11533} {\bibfield
  {journal} {\bibinfo  {journal} {Phys. Rev. B},\ }\textbf {\bibinfo {volume}
  {57}},\ \bibinfo {pages} {11533} (\bibinfo {year} {1998})}\BibitemShut
  {NoStop}%
\bibitem [{\citenamefont {Blumberg}\ \emph {et~al.}(2002)\citenamefont
  {Blumberg}, \citenamefont {Littlewood}, \citenamefont {Gozar}, \citenamefont
  {Dennis}, \citenamefont {Motoyama}, \citenamefont {Eisaki},\ and\
  \citenamefont {Uchida}}]{Blumberg2002}%
  \BibitemOpen
  \bibfield  {author} {\bibinfo {author} {\bibfnamefont {G.}~\bibnamefont
  {Blumberg}}, \bibinfo {author} {\bibfnamefont {P.}~\bibnamefont
  {Littlewood}}, \bibinfo {author} {\bibfnamefont {A.}~\bibnamefont {Gozar}},
  \bibinfo {author} {\bibfnamefont {B.}~\bibnamefont {Dennis}}, \bibinfo
  {author} {\bibfnamefont {N.}~\bibnamefont {Motoyama}}, \bibinfo {author}
  {\bibfnamefont {H.}~\bibnamefont {Eisaki}}, \ and\ \bibinfo {author}
  {\bibfnamefont {S.}~\bibnamefont {Uchida}},\ }\href@noop {} {\bibfield
  {journal} {\bibinfo  {journal} {Science},\ }\textbf {\bibinfo {volume}
  {297}},\ \bibinfo {pages} {584} (\bibinfo {year} {2002})}\BibitemShut
  {NoStop}%
\bibitem [{\citenamefont {Vuleti\'{c}}\ \emph {et~al.}(2003)\citenamefont
  {Vuleti\'{c}}, \citenamefont {Korin-Hamzi\'{c}}, \citenamefont {Tomi\'{c}},
  \citenamefont {Gorshunov}, \citenamefont {Haas}, \citenamefont {R\~o\ om},
  \citenamefont {Dressel}, \citenamefont {Akimitsu}, \citenamefont {Sasaki},\
  and\ \citenamefont {Nagata}}]{Vuletic2003}%
  \BibitemOpen
  \bibfield  {author} {\bibinfo {author} {\bibfnamefont {T.}~\bibnamefont
  {Vuleti\'{c}}}, \bibinfo {author} {\bibfnamefont {B.}~\bibnamefont
  {Korin-Hamzi\'{c}}}, \bibinfo {author} {\bibfnamefont {S.}~\bibnamefont
  {Tomi\'{c}}}, \bibinfo {author} {\bibfnamefont {B.}~\bibnamefont
  {Gorshunov}}, \bibinfo {author} {\bibfnamefont {P.}~\bibnamefont {Haas}},
  \bibinfo {author} {\bibfnamefont {T.}~\bibnamefont {R\~o\ om}}, \bibinfo
  {author} {\bibfnamefont {M.}~\bibnamefont {Dressel}}, \bibinfo {author}
  {\bibfnamefont {J.}~\bibnamefont {Akimitsu}}, \bibinfo {author}
  {\bibfnamefont {T.}~\bibnamefont {Sasaki}}, \ and\ \bibinfo {author}
  {\bibfnamefont {T.}~\bibnamefont {Nagata}},\ }\Doi
  {10.1103/PhysRevLett.90.257002} {\bibfield  {journal} {\bibinfo  {journal}
  {Phys. Rev. Lett.},\ }\textbf {\bibinfo {volume} {90}},\ \bibinfo {pages}
  {257002} (\bibinfo {year} {2003})}\BibitemShut {NoStop}%
\bibitem [{\citenamefont {Hess}\ \emph {et~al.}(2004)\citenamefont {Hess},
  \citenamefont {ElHaes}, \citenamefont {B\"uchner}, \citenamefont {Ammerahl},
  \citenamefont {H\"ucker},\ and\ \citenamefont {Revcolevschi}}]{Hess2004}%
  \BibitemOpen
  \bibfield  {author} {\bibinfo {author} {\bibfnamefont {C.}~\bibnamefont
  {Hess}}, \bibinfo {author} {\bibfnamefont {H.}~\bibnamefont {ElHaes}},
  \bibinfo {author} {\bibfnamefont {B.}~\bibnamefont {B\"uchner}}, \bibinfo
  {author} {\bibfnamefont {U.}~\bibnamefont {Ammerahl}}, \bibinfo {author}
  {\bibfnamefont {M.}~\bibnamefont {H\"ucker}}, \ and\ \bibinfo {author}
  {\bibfnamefont {A.}~\bibnamefont {Revcolevschi}},\ }\Doi
  {10.1103/PhysRevLett.93.027005} {\bibfield  {journal} {\bibinfo  {journal}
  {Phys. Rev. Lett.},\ }\textbf {\bibinfo {volume} {93}},\ \bibinfo {pages}
  {027005} (\bibinfo {year} {2004})}\BibitemShut {NoStop}%
\bibitem [{\citenamefont {Dagotto}\ \emph {et~al.}(1992)\citenamefont
  {Dagotto}, \citenamefont {Riera},\ and\ \citenamefont
  {Scalapino}}]{Dagotto1992}%
  \BibitemOpen
  \bibfield  {author} {\bibinfo {author} {\bibfnamefont {E.}~\bibnamefont
  {Dagotto}}, \bibinfo {author} {\bibfnamefont {J.}~\bibnamefont {Riera}}, \
  and\ \bibinfo {author} {\bibfnamefont {D.}~\bibnamefont {Scalapino}},\
  }\href@noop {} {\bibfield  {journal} {\bibinfo  {journal} {Phys. Rev. B},\
  }\textbf {\bibinfo {volume} {45}},\ \bibinfo {pages} {5744} (\bibinfo {year}
  {1992})}\BibitemShut {NoStop}%
\bibitem [{\citenamefont {Abbamonte}\ \emph {et~al.}(2004)\citenamefont
  {Abbamonte}, \citenamefont {Blumberg}, \citenamefont {Rusydi}, \citenamefont
  {Gozar}, \citenamefont {Evans}, \citenamefont {Siegrist}, \citenamefont
  {Venema}, \citenamefont {Eisaki}, \citenamefont {Isaacs},\ and\ \citenamefont
  {Sawatzky}}]{Abbamonte2004}%
  \BibitemOpen
  \bibfield  {author} {\bibinfo {author} {\bibfnamefont {P.}~\bibnamefont
  {Abbamonte}}, \bibinfo {author} {\bibfnamefont {G.}~\bibnamefont {Blumberg}},
  \bibinfo {author} {\bibfnamefont {A.}~\bibnamefont {Rusydi}}, \bibinfo
  {author} {\bibfnamefont {A.}~\bibnamefont {Gozar}}, \bibinfo {author}
  {\bibfnamefont {P.}~\bibnamefont {Evans}}, \bibinfo {author} {\bibfnamefont
  {T.}~\bibnamefont {Siegrist}}, \bibinfo {author} {\bibfnamefont
  {L.}~\bibnamefont {Venema}}, \bibinfo {author} {\bibfnamefont
  {H.}~\bibnamefont {Eisaki}}, \bibinfo {author} {\bibfnamefont
  {E.}~\bibnamefont {Isaacs}}, \ and\ \bibinfo {author} {\bibfnamefont
  {G.}~\bibnamefont {Sawatzky}},\ }\href@noop {} {\bibfield  {journal}
  {\bibinfo  {journal} {Nature},\ }\textbf {\bibinfo {volume} {431}},\ \bibinfo
  {pages} {1078} (\bibinfo {year} {2004})}\BibitemShut {NoStop}%
\bibitem [{\citenamefont {Carr}\ and\ \citenamefont
  {Tsvelik}(2002)}]{Carr2002}%
  \BibitemOpen
  \bibfield  {author} {\bibinfo {author} {\bibfnamefont {S.~T.}\ \bibnamefont
  {Carr}}\ and\ \bibinfo {author} {\bibfnamefont {A.~M.}\ \bibnamefont
  {Tsvelik}},\ }\Doi {10.1103/PhysRevB.65.195121} {\bibfield  {journal}
  {\bibinfo  {journal} {Phys. Rev. B},\ }\textbf {\bibinfo {volume} {65}},\
  \bibinfo {pages} {195121} (\bibinfo {year} {2002})}\BibitemShut {NoStop}%
\bibitem [{\citenamefont {White}\ \emph {et~al.}(2002)\citenamefont {White},
  \citenamefont {Affleck},\ and\ \citenamefont {Scalapino}}]{Friedel2002}%
  \BibitemOpen
  \bibfield  {author} {\bibinfo {author} {\bibfnamefont {S.~R.}\ \bibnamefont
  {White}}, \bibinfo {author} {\bibfnamefont {I.}~\bibnamefont {Affleck}}, \
  and\ \bibinfo {author} {\bibfnamefont {D.~J.}\ \bibnamefont {Scalapino}},\
  }\Doi {10.1103/PhysRevB.65.165122} {\bibfield  {journal} {\bibinfo  {journal}
  {Phys. Rev. B},\ }\textbf {\bibinfo {volume} {65}},\ \bibinfo {pages}
  {165122} (\bibinfo {year} {2002})}\BibitemShut {NoStop}%
\bibitem [{\citenamefont {Fink}\ \emph {et~al.}(2001)\citenamefont {Fink},
  \citenamefont {Knupfer}, \citenamefont {Atzkern},\ and\ \citenamefont
  {Golden}}]{Fink2001}%
  \BibitemOpen
  \bibfield  {author} {\bibinfo {author} {\bibfnamefont {J.}~\bibnamefont
  {Fink}}, \bibinfo {author} {\bibfnamefont {M.}~\bibnamefont {Knupfer}},
  \bibinfo {author} {\bibfnamefont {S.}~\bibnamefont {Atzkern}}, \ and\
  \bibinfo {author} {\bibfnamefont {M.}~\bibnamefont {Golden}},\ }\href@noop {}
  {\bibfield  {journal} {\bibinfo  {journal} {J. Electron Spectrosc. Relat.
  Phenom.},\ }\textbf {\bibinfo {volume} {117}},\ \bibinfo {pages} {287}
  (\bibinfo {year} {2001})}\BibitemShut {NoStop}%
\bibitem [{\citenamefont {Knupfer}\ and\ \citenamefont
  {Fink}(1999)}]{Knupfer1999}%
  \BibitemOpen
  \bibfield  {author} {\bibinfo {author} {\bibfnamefont {M.}~\bibnamefont
  {Knupfer}}\ and\ \bibinfo {author} {\bibfnamefont {J.}~\bibnamefont {Fink}},\
  }\Doi {10.1103/PhysRevB.60.10731} {\bibfield  {journal} {\bibinfo  {journal}
  {Phys. Rev. B},\ }\textbf {\bibinfo {volume} {60}},\ \bibinfo {pages} {10731}
  (\bibinfo {year} {1999})}\BibitemShut {NoStop}%
\bibitem [{\citenamefont {Atzkern}\ \emph {et~al.}(2000)\citenamefont
  {Atzkern}, \citenamefont {Knupfer}, \citenamefont {Golden}, \citenamefont
  {Fink}, \citenamefont {Waidacher}, \citenamefont {Richter}, \citenamefont
  {Becker}, \citenamefont {Motoyama}, \citenamefont {Eisaki},\ and\
  \citenamefont {Uchida}}]{Atzkern2000}%
  \BibitemOpen
  \bibfield  {author} {\bibinfo {author} {\bibfnamefont {S.}~\bibnamefont
  {Atzkern}}, \bibinfo {author} {\bibfnamefont {M.}~\bibnamefont {Knupfer}},
  \bibinfo {author} {\bibfnamefont {M.~S.}\ \bibnamefont {Golden}}, \bibinfo
  {author} {\bibfnamefont {J.}~\bibnamefont {Fink}}, \bibinfo {author}
  {\bibfnamefont {C.}~\bibnamefont {Waidacher}}, \bibinfo {author}
  {\bibfnamefont {J.}~\bibnamefont {Richter}}, \bibinfo {author} {\bibfnamefont
  {K.~W.}\ \bibnamefont {Becker}}, \bibinfo {author} {\bibfnamefont
  {N.}~\bibnamefont {Motoyama}}, \bibinfo {author} {\bibfnamefont
  {H.}~\bibnamefont {Eisaki}}, \ and\ \bibinfo {author} {\bibfnamefont
  {S.}~\bibnamefont {Uchida}},\ }\Doi {10.1103/PhysRevB.62.7845} {\bibfield
  {journal} {\bibinfo  {journal} {Phys. Rev. B},\ }\textbf {\bibinfo {volume}
  {62}},\ \bibinfo {pages} {7845} (\bibinfo {year} {2000})}\BibitemShut
  {NoStop}%
\bibitem [{\citenamefont {N{\"u}cker}\ \emph {et~al.}(1989)\citenamefont
  {N{\"u}cker}, \citenamefont {Romberg}, \citenamefont {Nakai}, \citenamefont
  {Scheerer}, \citenamefont {Fink}, \citenamefont {Yan},\ and\ \citenamefont
  {Zhao}}]{Nuecker1989}%
  \BibitemOpen
  \bibfield  {author} {\bibinfo {author} {\bibfnamefont {N.}~\bibnamefont
  {N{\"u}cker}}, \bibinfo {author} {\bibfnamefont {H.}~\bibnamefont {Romberg}},
  \bibinfo {author} {\bibfnamefont {S.}~\bibnamefont {Nakai}}, \bibinfo
  {author} {\bibfnamefont {B.}~\bibnamefont {Scheerer}}, \bibinfo {author}
  {\bibfnamefont {J.}~\bibnamefont {Fink}}, \bibinfo {author} {\bibfnamefont
  {Y.~F.}\ \bibnamefont {Yan}}, \ and\ \bibinfo {author} {\bibfnamefont
  {Z.~X.}\ \bibnamefont {Zhao}},\ }\Doi {10.1103/PhysRevB.39.12379} {\bibfield
  {journal} {\bibinfo  {journal} {Phys. Rev. B},\ }\textbf {\bibinfo {volume}
  {39}},\ \bibinfo {pages} {12379} (\bibinfo {year} {1989})}\BibitemShut
  {NoStop}%
\bibitem [{\citenamefont {Wang}\ \emph {et~al.}(1990)\citenamefont {Wang},
  \citenamefont {Feng},\ and\ \citenamefont {Ritter}}]{Wang1990}%
  \BibitemOpen
  \bibfield  {author} {\bibinfo {author} {\bibfnamefont {Y.-Y.}\ \bibnamefont
  {Wang}}, \bibinfo {author} {\bibfnamefont {G.}~\bibnamefont {Feng}}, \ and\
  \bibinfo {author} {\bibfnamefont {A.~L.}\ \bibnamefont {Ritter}},\ }\Doi
  {10.1103/PhysRevB.42.420} {\bibfield  {journal} {\bibinfo  {journal} {Phys.
  Rev. B},\ }\textbf {\bibinfo {volume} {42}},\ \bibinfo {pages} {420}
  (\bibinfo {year} {1990})}\BibitemShut {NoStop}%
\bibitem [{\citenamefont {N{\"u}cker}\ \emph {et~al.}(1991)\citenamefont
  {N{\"u}cker}, \citenamefont {Eckern}, \citenamefont {Fink},\ and\
  \citenamefont {M{\"u}ller}}]{Nuecker1991}%
  \BibitemOpen
  \bibfield  {author} {\bibinfo {author} {\bibfnamefont {N.}~\bibnamefont
  {N{\"u}cker}}, \bibinfo {author} {\bibfnamefont {U.}~\bibnamefont {Eckern}},
  \bibinfo {author} {\bibfnamefont {J.}~\bibnamefont {Fink}}, \ and\ \bibinfo
  {author} {\bibfnamefont {P.}~\bibnamefont {M{\"u}ller}},\ }\Doi
  {10.1103/PhysRevB.44.7155} {\bibfield  {journal} {\bibinfo  {journal} {Phys.
  Rev. B},\ }\textbf {\bibinfo {volume} {44}},\ \bibinfo {pages} {7155}
  (\bibinfo {year} {1991})}\BibitemShut {NoStop}%
\bibitem [{\citenamefont {Romberg}\ \emph {et~al.}(1990)\citenamefont
  {Romberg}, \citenamefont {N{\"u}cker}, \citenamefont {Fink}, \citenamefont
  {Wolf}, \citenamefont {Xi}, \citenamefont {Koch}, \citenamefont {Geserich},
  \citenamefont {Durrler}, \citenamefont {Assmus},\ and\ \citenamefont
  {Gegenheimer}}]{Romberg1990}%
  \BibitemOpen
  \bibfield  {author} {\bibinfo {author} {\bibfnamefont {H.}~\bibnamefont
  {Romberg}}, \bibinfo {author} {\bibfnamefont {N.}~\bibnamefont {N{\"u}cker}},
  \bibinfo {author} {\bibfnamefont {J.}~\bibnamefont {Fink}}, \bibinfo {author}
  {\bibfnamefont {T.}~\bibnamefont {Wolf}}, \bibinfo {author} {\bibfnamefont
  {X.}~\bibnamefont {Xi}}, \bibinfo {author} {\bibfnamefont {B.}~\bibnamefont
  {Koch}}, \bibinfo {author} {\bibfnamefont {H.}~\bibnamefont {Geserich}},
  \bibinfo {author} {\bibfnamefont {M.}~\bibnamefont {Durrler}}, \bibinfo
  {author} {\bibfnamefont {W.}~\bibnamefont {Assmus}}, \ and\ \bibinfo {author}
  {\bibfnamefont {B.}~\bibnamefont {Gegenheimer}},\ }\href@noop {} {\bibfield
  {journal} {\bibinfo  {journal} {Z. Phys. B-Condens. Mat.},\ }\textbf
  {\bibinfo {volume} {78}},\ \bibinfo {pages} {367} (\bibinfo {year}
  {1990})}\BibitemShut {NoStop}%
\bibitem [{\citenamefont {Schuster}\ \emph {et~al.}(2007)\citenamefont
  {Schuster}, \citenamefont {Knupfer},\ and\ \citenamefont
  {Berger}}]{Schuster2007}%
  \BibitemOpen
  \bibfield  {author} {\bibinfo {author} {\bibfnamefont {R.}~\bibnamefont
  {Schuster}}, \bibinfo {author} {\bibfnamefont {M.}~\bibnamefont {Knupfer}}, \
  and\ \bibinfo {author} {\bibfnamefont {H.}~\bibnamefont {Berger}},\ }\Doi
  {10.1103/PhysRevLett.98.037402} {\bibfield  {journal} {\bibinfo  {journal}
  {Phys. Rev. Lett.},\ }\textbf {\bibinfo {volume} {98}},\ \bibinfo {pages}
  {037402} (\bibinfo {year} {2007})}\BibitemShut {NoStop}%
\bibitem [{\citenamefont {Ammerahl}\ \emph {et~al.}(1998)\citenamefont
  {Ammerahl}, \citenamefont {Dhalenne}, \citenamefont {Revcolevschi},
  \citenamefont {Berthon},\ and\ \citenamefont {Moudden}}]{Ammerahl1998}%
  \BibitemOpen
  \bibfield  {author} {\bibinfo {author} {\bibfnamefont {U.}~\bibnamefont
  {Ammerahl}}, \bibinfo {author} {\bibfnamefont {G.}~\bibnamefont {Dhalenne}},
  \bibinfo {author} {\bibfnamefont {A.}~\bibnamefont {Revcolevschi}}, \bibinfo
  {author} {\bibfnamefont {J.}~\bibnamefont {Berthon}}, \ and\ \bibinfo
  {author} {\bibfnamefont {H.}~\bibnamefont {Moudden}},\ }\Doi {DOI:
  10.1016/S0022-0248(98)00487-4} {\bibfield  {journal} {\bibinfo  {journal}
  {Journal of Crystal Growth},\ }\textbf {\bibinfo {volume} {193}},\ \bibinfo
  {pages} {55} (\bibinfo {year} {1998})}\BibitemShut {NoStop}%
\bibitem [{\citenamefont {Fink}(1989)}]{Fink1989}%
  \BibitemOpen
  \bibfield  {author} {\bibinfo {author} {\bibfnamefont {J.}~\bibnamefont
  {Fink}},\ }\href@noop {} {\bibfield  {journal} {\bibinfo  {journal} {Adv.
  Electron. Electron Phys.},\ }\textbf {\bibinfo {volume} {75}},\ \bibinfo
  {pages} {121} (\bibinfo {year} {1989})}\BibitemShut {NoStop}%
\bibitem [{\citenamefont {Chen}\ and\ \citenamefont {Silcox}(1977)}]{Chen1977}%
  \BibitemOpen
  \bibfield  {author} {\bibinfo {author} {\bibfnamefont {C.~H.}\ \bibnamefont
  {Chen}}\ and\ \bibinfo {author} {\bibfnamefont {J.}~\bibnamefont {Silcox}},\
  }\Doi {10.1103/PhysRevB.16.4246} {\bibfield  {journal} {\bibinfo  {journal}
  {Phys. Rev. B},\ }\textbf {\bibinfo {volume} {16}},\ \bibinfo {pages} {4246}
  (\bibinfo {year} {1977})}\BibitemShut {NoStop}%
\bibitem [{\citenamefont {Batson}\ and\ \citenamefont
  {Silcox}(1983)}]{Batson1983}%
  \BibitemOpen
  \bibfield  {author} {\bibinfo {author} {\bibfnamefont {P.~E.}\ \bibnamefont
  {Batson}}\ and\ \bibinfo {author} {\bibfnamefont {J.}~\bibnamefont
  {Silcox}},\ }\Doi {10.1103/PhysRevB.27.5224} {\bibfield  {journal} {\bibinfo
  {journal} {Phys. Rev. B},\ }\textbf {\bibinfo {volume} {27}},\ \bibinfo
  {pages} {5224} (\bibinfo {year} {1983})}\BibitemShut {NoStop}%
\bibitem [{\citenamefont {Bertoni}\ and\ \citenamefont
  {Brosens}()}]{Bertoni2010}%
  \BibitemOpen
  \bibfield  {author} {\bibinfo {author} {\bibfnamefont {V.~J.}\ \bibnamefont
  {Bertoni}, \bibfnamefont {G.}}\ and\ \bibinfo {author} {\bibfnamefont
  {F.}~\bibnamefont {Brosens}},\ }\Doi {10.1002/jemt.20868} {\bibfield
  {journal} {\bibinfo  {journal} {Micr. Res. Tech.}},\ }\doi
  {10.1002/jemt.20868}\BibitemShut {NoStop}%
\bibitem [{\citenamefont {Motoyama}\ \emph {et~al.}(1997)\citenamefont
  {Motoyama}, \citenamefont {Osafune}, \citenamefont {Kakeshita}, \citenamefont
  {Eisaki},\ and\ \citenamefont {Uchida}}]{Motoyama1997}%
  \BibitemOpen
  \bibfield  {author} {\bibinfo {author} {\bibfnamefont {N.}~\bibnamefont
  {Motoyama}}, \bibinfo {author} {\bibfnamefont {T.}~\bibnamefont {Osafune}},
  \bibinfo {author} {\bibfnamefont {T.}~\bibnamefont {Kakeshita}}, \bibinfo
  {author} {\bibfnamefont {H.}~\bibnamefont {Eisaki}}, \ and\ \bibinfo {author}
  {\bibfnamefont {S.}~\bibnamefont {Uchida}},\ }\Doi
  {10.1103/PhysRevB.55.R3386} {\bibfield  {journal} {\bibinfo  {journal} {Phys.
  Rev. B},\ }\textbf {\bibinfo {volume} {55}},\ \bibinfo {pages} {R3386}
  (\bibinfo {year} {1997})}\BibitemShut {NoStop}%
\bibitem [{\citenamefont {Ruzicka}\ \emph {et~al.}(1998)\citenamefont
  {Ruzicka}, \citenamefont {Degiorgi}, \citenamefont {Ammerahl}, \citenamefont
  {Dhalenne},\ and\ \citenamefont {Revcolevschi}}]{Ruzicka1998}%
  \BibitemOpen
  \bibfield  {author} {\bibinfo {author} {\bibfnamefont {B.}~\bibnamefont
  {Ruzicka}}, \bibinfo {author} {\bibfnamefont {L.}~\bibnamefont {Degiorgi}},
  \bibinfo {author} {\bibfnamefont {U.}~\bibnamefont {Ammerahl}}, \bibinfo
  {author} {\bibfnamefont {G.}~\bibnamefont {Dhalenne}}, \ and\ \bibinfo
  {author} {\bibfnamefont {A.}~\bibnamefont {Revcolevschi}},\ }\href@noop {}
  {\bibfield  {journal} {\bibinfo  {journal} {Eur. Phys. J. B},\ }\textbf
  {\bibinfo {volume} {6}},\ \bibinfo {pages} {301} (\bibinfo {year}
  {1998})}\BibitemShut {NoStop}%
\bibitem [{\citenamefont {Knupfer}\ \emph {et~al.}(1994)\citenamefont
  {Knupfer}, \citenamefont {Roth}, \citenamefont {Fink}, \citenamefont
  {Karpinski},\ and\ \citenamefont {Kaldis}}]{Knupfer1994}%
  \BibitemOpen
  \bibfield  {author} {\bibinfo {author} {\bibfnamefont {M.}~\bibnamefont
  {Knupfer}}, \bibinfo {author} {\bibfnamefont {G.}~\bibnamefont {Roth}},
  \bibinfo {author} {\bibfnamefont {J.}~\bibnamefont {Fink}}, \bibinfo {author}
  {\bibfnamefont {J.}~\bibnamefont {Karpinski}}, \ and\ \bibinfo {author}
  {\bibfnamefont {E.}~\bibnamefont {Kaldis}},\ }\Doi {DOI:
  10.1016/0921-4534(94)90453-7} {\bibfield  {journal} {\bibinfo  {journal}
  {Physica C: Superconductivity},\ }\textbf {\bibinfo {volume} {230}},\
  \bibinfo {pages} {121} (\bibinfo {year} {1994})}\BibitemShut {NoStop}%
\bibitem [{\citenamefont {Sturm}(1982)}]{Sturm1982}%
  \BibitemOpen
  \bibfield  {author} {\bibinfo {author} {\bibfnamefont {K.}~\bibnamefont
  {Sturm}},\ }\href@noop {} {\bibfield  {journal} {\bibinfo  {journal} {Adv.
  Phys.},\ }\textbf {\bibinfo {volume} {31}},\ \bibinfo {pages} {1} (\bibinfo
  {year} {1982})}\BibitemShut {NoStop}%
\bibitem [{\citenamefont {Pines}(1963)}]{Pines1963}%
  \BibitemOpen
  \bibfield  {author} {\bibinfo {author} {\bibfnamefont {D.}~\bibnamefont
  {Pines}},\ }\href@noop {} {\emph {\bibinfo {title} {Elementary Excitations In
  Solids}}}\ (\bibinfo  {publisher} {W. A. Benjamin, Inc. N.Y.},\ \bibinfo
  {year} {1963})\BibitemShut {NoStop}%
\bibitem [{\citenamefont {Williams}\ and\ \citenamefont
  {Bloch}(1974)}]{Williams1974}%
  \BibitemOpen
  \bibfield  {author} {\bibinfo {author} {\bibfnamefont {P.~F.}\ \bibnamefont
  {Williams}}\ and\ \bibinfo {author} {\bibfnamefont {A.~N.}\ \bibnamefont
  {Bloch}},\ }\Doi {10.1103/PhysRevB.10.1097} {\bibfield  {journal} {\bibinfo
  {journal} {Phys. Rev. B},\ }\textbf {\bibinfo {volume} {10}},\ \bibinfo
  {pages} {1097} (\bibinfo {year} {1974})}\BibitemShut {NoStop}%
\bibitem [{\citenamefont {Kojima}\ \emph {et~al.}(2001)\citenamefont {Kojima},
  \citenamefont {Motoyama}, \citenamefont {Eisaki},\ and\ \citenamefont
  {Uchida}}]{Kojima2001}%
  \BibitemOpen
  \bibfield  {author} {\bibinfo {author} {\bibfnamefont {K.~M.}\ \bibnamefont
  {Kojima}}, \bibinfo {author} {\bibfnamefont {N.}~\bibnamefont {Motoyama}},
  \bibinfo {author} {\bibfnamefont {H.}~\bibnamefont {Eisaki}}, \ and\ \bibinfo
  {author} {\bibfnamefont {S.}~\bibnamefont {Uchida}},\ }\Doi {DOI:
  10.1016/S0368-2048(01)00268-7} {\bibfield  {journal} {\bibinfo  {journal} {J.
  Electron Spectrosc. Relat. Phenom.},\ }\textbf {\bibinfo {volume}
  {117-118}},\ \bibinfo {pages} {237 } (\bibinfo {year} {2001})}\BibitemShut
  {NoStop}%
\bibitem [{\citenamefont {Hu}\ \emph {et~al.}(2002)\citenamefont {Hu},
  \citenamefont {Drechsler}, \citenamefont {Malek}, \citenamefont {Rosner},
  \citenamefont {Neudert}, \citenamefont {Knupfer}, \citenamefont {Golden},
  \citenamefont {Fink}, \citenamefont {Karpinski}, \citenamefont {Kaindl},
  \citenamefont {Hellwig},\ and\ \citenamefont {Jung}}]{Hu2002}%
  \BibitemOpen
  \bibfield  {author} {\bibinfo {author} {\bibfnamefont {Z.}~\bibnamefont
  {Hu}}, \bibinfo {author} {\bibfnamefont {S.}~\bibnamefont {Drechsler}},
  \bibinfo {author} {\bibfnamefont {J.}~\bibnamefont {Malek}}, \bibinfo
  {author} {\bibfnamefont {H.}~\bibnamefont {Rosner}}, \bibinfo {author}
  {\bibfnamefont {R.}~\bibnamefont {Neudert}}, \bibinfo {author} {\bibfnamefont
  {M.}~\bibnamefont {Knupfer}}, \bibinfo {author} {\bibfnamefont
  {M.}~\bibnamefont {Golden}}, \bibinfo {author} {\bibfnamefont
  {J.}~\bibnamefont {Fink}}, \bibinfo {author} {\bibfnamefont {J.}~\bibnamefont
  {Karpinski}}, \bibinfo {author} {\bibfnamefont {G.}~\bibnamefont {Kaindl}},
  \bibinfo {author} {\bibfnamefont {C.}~\bibnamefont {Hellwig}}, \ and\
  \bibinfo {author} {\bibfnamefont {C.}~\bibnamefont {Jung}},\ }\href@noop {}
  {\bibfield  {journal} {\bibinfo  {journal} {Europhys. Lett.},\ }\textbf
  {\bibinfo {volume} {59}},\ \bibinfo {pages} {135} (\bibinfo {year}
  {2002})}\BibitemShut {NoStop}%
\bibitem [{\citenamefont {Zhang}\ and\ \citenamefont {Ng}(1998)}]{Zhang1998}%
  \BibitemOpen
  \bibfield  {author} {\bibinfo {author} {\bibfnamefont {F.~C.}\ \bibnamefont
  {Zhang}}\ and\ \bibinfo {author} {\bibfnamefont {K.~K.}\ \bibnamefont {Ng}},\
  }\Doi {10.1103/PhysRevB.58.13520} {\bibfield  {journal} {\bibinfo  {journal}
  {Phys. Rev. B},\ }\textbf {\bibinfo {volume} {58}},\ \bibinfo {pages} {13520}
  (\bibinfo {year} {1998})}\BibitemShut {NoStop}%
\bibitem [{\citenamefont {Wang}\ \emph {et~al.}(1996)\citenamefont {Wang},
  \citenamefont {Zhang}, \citenamefont {Dravid}, \citenamefont {Ng},
  \citenamefont {Klein}, \citenamefont {Schnatterly},\ and\ \citenamefont
  {Miller}}]{Wang1996}%
  \BibitemOpen
  \bibfield  {author} {\bibinfo {author} {\bibfnamefont {Y.~Y.}\ \bibnamefont
  {Wang}}, \bibinfo {author} {\bibfnamefont {F.~C.}\ \bibnamefont {Zhang}},
  \bibinfo {author} {\bibfnamefont {V.~P.}\ \bibnamefont {Dravid}}, \bibinfo
  {author} {\bibfnamefont {K.~K.}\ \bibnamefont {Ng}}, \bibinfo {author}
  {\bibfnamefont {M.~V.}\ \bibnamefont {Klein}}, \bibinfo {author}
  {\bibfnamefont {S.~E.}\ \bibnamefont {Schnatterly}}, \ and\ \bibinfo {author}
  {\bibfnamefont {L.~L.}\ \bibnamefont {Miller}},\ }\Doi
  {10.1103/PhysRevLett.77.1809} {\bibfield  {journal} {\bibinfo  {journal}
  {Phys. Rev. Lett.},\ }\textbf {\bibinfo {volume} {77}},\ \bibinfo {pages}
  {1809} (\bibinfo {year} {1996})}\BibitemShut {NoStop}%
\bibitem [{\citenamefont {Zhang}\ and\ \citenamefont {Rice}(1988)}]{Zhang1988}%
  \BibitemOpen
  \bibfield  {author} {\bibinfo {author} {\bibfnamefont {F.~C.}\ \bibnamefont
  {Zhang}}\ and\ \bibinfo {author} {\bibfnamefont {T.~M.}\ \bibnamefont
  {Rice}},\ }\Doi {10.1103/PhysRevB.37.3759} {\bibfield  {journal} {\bibinfo
  {journal} {Phys. Rev. B},\ }\textbf {\bibinfo {volume} {37}},\ \bibinfo
  {pages} {3759} (\bibinfo {year} {1988})}\BibitemShut {NoStop}%
\bibitem [{\citenamefont {Schuster}(2010)}]{Schuster2010}%
  \BibitemOpen
  \bibfield  {author} {\bibinfo {author} {\bibfnamefont {R.}~\bibnamefont
  {Schuster}},\ }\emph {\bibinfo {title} {Electron Energy-Loss Spectroscopy On
  Underdoped Cuprates And Transition-Metal Dichalcogenides}},\ \href@noop {}
  {Ph.D. thesis},\ \bibinfo  {school} {TU Dresden} (\bibinfo {year}
  {2010})\BibitemShut {NoStop}%
\bibitem [{\citenamefont {Wray}\ \emph
  {et~al.}(2008){\natexlab{a}}\citenamefont {Wray}, \citenamefont {Qian},
  \citenamefont {Hsieh}, \citenamefont {Xia}, \citenamefont {Gog},
  \citenamefont {Casa}, \citenamefont {Eisaki},\ and\ \citenamefont
  {Hasan}}]{Wray2008}%
  \BibitemOpen
  \bibfield  {author} {\bibinfo {author} {\bibfnamefont {L.}~\bibnamefont
  {Wray}}, \bibinfo {author} {\bibfnamefont {D.}~\bibnamefont {Qian}}, \bibinfo
  {author} {\bibfnamefont {D.}~\bibnamefont {Hsieh}}, \bibinfo {author}
  {\bibfnamefont {Y.}~\bibnamefont {Xia}}, \bibinfo {author} {\bibfnamefont
  {T.}~\bibnamefont {Gog}}, \bibinfo {author} {\bibfnamefont {D.}~\bibnamefont
  {Casa}}, \bibinfo {author} {\bibfnamefont {H.}~\bibnamefont {Eisaki}}, \ and\
  \bibinfo {author} {\bibfnamefont {M.}~\bibnamefont {Hasan}},\ }\Doi {DOI:
  10.1016/j.physb.2007.10.302} {\bibfield  {journal} {\bibinfo  {journal}
  {Physica B},\ }\textbf {\bibinfo {volume} {403}},\ \bibinfo {pages} {1456 }
  (\bibinfo {year} {2008}{\natexlab{a}})}\BibitemShut {NoStop}%
\bibitem [{\citenamefont {Wray}\ \emph
  {et~al.}(2008){\natexlab{b}}\citenamefont {Wray}, \citenamefont {Qian},
  \citenamefont {Hsieh}, \citenamefont {Xia}, \citenamefont {Gog},
  \citenamefont {Casa}, \citenamefont {Eisaki},\ and\ \citenamefont
  {Hasan}}]{Wray2008_2}%
  \BibitemOpen
  \bibfield  {author} {\bibinfo {author} {\bibfnamefont {L.}~\bibnamefont
  {Wray}}, \bibinfo {author} {\bibfnamefont {D.}~\bibnamefont {Qian}}, \bibinfo
  {author} {\bibfnamefont {D.}~\bibnamefont {Hsieh}}, \bibinfo {author}
  {\bibfnamefont {Y.}~\bibnamefont {Xia}}, \bibinfo {author} {\bibfnamefont
  {T.}~\bibnamefont {Gog}}, \bibinfo {author} {\bibfnamefont {D.}~\bibnamefont
  {Casa}}, \bibinfo {author} {\bibfnamefont {H.}~\bibnamefont {Eisaki}}, \ and\
  \bibinfo {author} {\bibfnamefont {M.~Z.}\ \bibnamefont {Hasan}},\ }\Doi
  {10.1016/j.jpcs.2008.06.112} {\bibfield  {journal} {\bibinfo  {journal} {J.
  Phys. Chem. Solids},\ }\textbf {\bibinfo {volume} {69}},\ \bibinfo {pages}
  {3146} (\bibinfo {year} {2008}{\natexlab{b}})}\BibitemShut {NoStop}%
\bibitem [{\citenamefont {Ishii}\ \emph {et~al.}(2007)\citenamefont {Ishii},
  \citenamefont {Tsutsui}, \citenamefont {Tohyama}, \citenamefont {Inami},
  \citenamefont {Mizuki}, \citenamefont {Murakami}, \citenamefont {Endoh},
  \citenamefont {Maekawa}, \citenamefont {Kudo}, \citenamefont {Koike},\ and\
  \citenamefont {Kumagai}}]{Ishii2007}%
  \BibitemOpen
  \bibfield  {author} {\bibinfo {author} {\bibfnamefont {K.}~\bibnamefont
  {Ishii}}, \bibinfo {author} {\bibfnamefont {K.}~\bibnamefont {Tsutsui}},
  \bibinfo {author} {\bibfnamefont {T.}~\bibnamefont {Tohyama}}, \bibinfo
  {author} {\bibfnamefont {T.}~\bibnamefont {Inami}}, \bibinfo {author}
  {\bibfnamefont {J.}~\bibnamefont {Mizuki}}, \bibinfo {author} {\bibfnamefont
  {Y.}~\bibnamefont {Murakami}}, \bibinfo {author} {\bibfnamefont
  {Y.}~\bibnamefont {Endoh}}, \bibinfo {author} {\bibfnamefont
  {S.}~\bibnamefont {Maekawa}}, \bibinfo {author} {\bibfnamefont
  {K.}~\bibnamefont {Kudo}}, \bibinfo {author} {\bibfnamefont {Y.}~\bibnamefont
  {Koike}}, \ and\ \bibinfo {author} {\bibfnamefont {K.}~\bibnamefont
  {Kumagai}},\ }\Doi {10.1103/PhysRevB.76.045124} {\bibfield  {journal}
  {\bibinfo  {journal} {Phys. Rev. B},\ }\textbf {\bibinfo {volume} {76}},\
  \bibinfo {pages} {045124} (\bibinfo {year} {2007})}\BibitemShut {NoStop}%
\bibitem [{\citenamefont {Sing}\ \emph
  {et~al.}(1999){\natexlab{a}}\citenamefont {Sing}, \citenamefont {Grigoryan},
  \citenamefont {Paasch}, \citenamefont {Knupfer}, \citenamefont {Fink},
  \citenamefont {Levy}, \citenamefont {Berger}, \citenamefont {Lommel},\ and\
  \citenamefont {Assmus}}]{Sing1999}%
  \BibitemOpen
  \bibfield  {author} {\bibinfo {author} {\bibfnamefont {M.}~\bibnamefont
  {Sing}}, \bibinfo {author} {\bibfnamefont {V.}~\bibnamefont {Grigoryan}},
  \bibinfo {author} {\bibfnamefont {G.}~\bibnamefont {Paasch}}, \bibinfo
  {author} {\bibfnamefont {M.}~\bibnamefont {Knupfer}}, \bibinfo {author}
  {\bibfnamefont {J.}~\bibnamefont {Fink}}, \bibinfo {author} {\bibfnamefont
  {F.}~\bibnamefont {Levy}}, \bibinfo {author} {\bibfnamefont {H.}~\bibnamefont
  {Berger}}, \bibinfo {author} {\bibfnamefont {B.}~\bibnamefont {Lommel}}, \
  and\ \bibinfo {author} {\bibfnamefont {W.}~\bibnamefont {Assmus}},\
  }\href@noop {} {\bibfield  {journal} {\bibinfo  {journal} {Synthetic
  Metals},\ }\textbf {\bibinfo {volume} {102}},\ \bibinfo {pages} {1591}
  (\bibinfo {year} {1999}{\natexlab{a}})}\BibitemShut {NoStop}%
\bibitem [{\citenamefont {Sing}(1999)}]{SingPhD}%
  \BibitemOpen
  \bibfield  {author} {\bibinfo {author} {\bibfnamefont {M.}~\bibnamefont
  {Sing}},\ }\emph {\bibinfo {title} {Zur Plasmonendispersion in
  quasi-eindimensionalen Leitern}},\ \href@noop {} {Ph.D. thesis},\ \bibinfo
  {school} {TU Dresden} (\bibinfo {year} {1999})\BibitemShut {NoStop}%
\bibitem [{\citenamefont {Arai}\ and\ \citenamefont
  {Tsunetsugu}(1997)}]{Arai1997}%
  \BibitemOpen
  \bibfield  {author} {\bibinfo {author} {\bibfnamefont {M.}~\bibnamefont
  {Arai}}\ and\ \bibinfo {author} {\bibfnamefont {H.}~\bibnamefont
  {Tsunetsugu}},\ }\href@noop {} {\bibfield  {journal} {\bibinfo  {journal}
  {Phys. Rev. B},\ }\textbf {\bibinfo {volume} {56}},\ \bibinfo {pages} {R4305}
  (\bibinfo {year} {1997})}\BibitemShut {NoStop}%
\bibitem [{\citenamefont {Sing}\ \emph
  {et~al.}(1999){\natexlab{b}}\citenamefont {Sing}, \citenamefont {Grigoryan},
  \citenamefont {Paasch}, \citenamefont {Knupfer}, \citenamefont {Fink},
  \citenamefont {Lommel},\ and\ \citenamefont {A\ss{}mus}}]{Sing1999_2}%
  \BibitemOpen
  \bibfield  {author} {\bibinfo {author} {\bibfnamefont {M.}~\bibnamefont
  {Sing}}, \bibinfo {author} {\bibfnamefont {V.~G.}\ \bibnamefont {Grigoryan}},
  \bibinfo {author} {\bibfnamefont {G.}~\bibnamefont {Paasch}}, \bibinfo
  {author} {\bibfnamefont {M.}~\bibnamefont {Knupfer}}, \bibinfo {author}
  {\bibfnamefont {J.}~\bibnamefont {Fink}}, \bibinfo {author} {\bibfnamefont
  {B.}~\bibnamefont {Lommel}}, \ and\ \bibinfo {author} {\bibfnamefont
  {W.}~\bibnamefont {A\ss{}mus}},\ }\Doi {10.1103/PhysRevB.59.5414} {\bibfield
  {journal} {\bibinfo  {journal} {Phys. Rev. B},\ }\textbf {\bibinfo {volume}
  {59}},\ \bibinfo {pages} {5414} (\bibinfo {year}
  {1999}{\natexlab{b}})}\BibitemShut {NoStop}%
\bibitem [{\citenamefont {vom Felde}\ \emph {et~al.}(1989)\citenamefont {vom
  Felde}, \citenamefont {Spr\"osser-Prou},\ and\ \citenamefont
  {Fink}}]{Felde1989}%
  \BibitemOpen
  \bibfield  {author} {\bibinfo {author} {\bibfnamefont {A.}~\bibnamefont {vom
  Felde}}, \bibinfo {author} {\bibfnamefont {J.}~\bibnamefont
  {Spr\"osser-Prou}}, \ and\ \bibinfo {author} {\bibfnamefont {J.}~\bibnamefont
  {Fink}},\ }\Doi {10.1103/PhysRevB.40.10181} {\bibfield  {journal} {\bibinfo
  {journal} {Phys. Rev. B},\ }\textbf {\bibinfo {volume} {40}},\ \bibinfo
  {pages} {10181} (\bibinfo {year} {1989})}\BibitemShut {NoStop}%
\bibitem [{\citenamefont {Sing}\ \emph {et~al.}(1998)\citenamefont {Sing},
  \citenamefont {Grigoryan}, \citenamefont {Paasch}, \citenamefont {Knupfer},
  \citenamefont {Fink}, \citenamefont {Berger},\ and\ \citenamefont
  {L\'evy}}]{Sing1998}%
  \BibitemOpen
  \bibfield  {author} {\bibinfo {author} {\bibfnamefont {M.}~\bibnamefont
  {Sing}}, \bibinfo {author} {\bibfnamefont {V.~G.}\ \bibnamefont {Grigoryan}},
  \bibinfo {author} {\bibfnamefont {G.}~\bibnamefont {Paasch}}, \bibinfo
  {author} {\bibfnamefont {M.}~\bibnamefont {Knupfer}}, \bibinfo {author}
  {\bibfnamefont {J.}~\bibnamefont {Fink}}, \bibinfo {author} {\bibfnamefont
  {H.}~\bibnamefont {Berger}}, \ and\ \bibinfo {author} {\bibfnamefont
  {F.}~\bibnamefont {L\'evy}},\ }\Doi {10.1103/PhysRevB.57.12768} {\bibfield
  {journal} {\bibinfo  {journal} {Phys. Rev. B},\ }\textbf {\bibinfo {volume}
  {57}},\ \bibinfo {pages} {12768} (\bibinfo {year} {1998})}\BibitemShut
  {NoStop}%
\end{thebibliography}%
\bibliographystyle{apsrev4-1}
\end{document}